\documentclass[longauth]{aa} 
\usepackage[switch]{lineno}
\usepackage{graphicx}
\usepackage{mwe}
\usepackage{amsmath}
\usepackage{tabularx}
\usepackage{multirow}
\usepackage{natbib}
\usepackage{xcolor}
\usepackage{bbm}
\usepackage{array}
\usepackage{lscape} 
\usepackage{ragged2e}
\DeclareUnicodeCharacter{0308}{HERE!HERE!}
\usepackage{graphicx}
\usepackage{txfonts}
\usepackage{booktabs}
\usepackage{hyperref}
\hypersetup{
    colorlinks=false,
    linkcolor=red,
    citecolor=blue,
    urlcolor=blue,
    filecolor=magenta, 
    pdfborder={1 1 1},
}

\begin{document}

   \title{{\sc{CircleZ}} : Reliable Photometric redshifts for AGN computed using only photometry from Legacy Survey Imaging for DESI} 
   

\author{A. Saxena\fnmsep\thanks{mara@mpe.mpg.de}\inst{1,2},
M. Salvato\inst{1,3},
W. Roster\inst{1},
R. Shirley\inst{1},
J. Buchner\inst{1},
J. Wolf\inst{1,3},
C. Kohl\inst{1},
H. Starck\inst{1},
T. Dwelly\inst{1},
J. Comparat\inst{1},
A. Malyali\inst{1},
S. Krippendorf\inst{4,5},
A. Zenteno\inst{6},
D. Lang\inst{7},
D. Schlegel\inst{8},
R. Zhou\inst{8},
A. Dey\inst{9},
F. Valdes\inst{10},
A. Myers\inst{11},
R. J. Assef\inst{12},
C. Ricci\inst{13,14}, 
M. J. Temple\inst{15},
A. Merloni\inst{1},
A. Koekemoer\inst{16},
S.F. Anderson\inst{17},
S. Morrison\inst{18},
X. Liu\inst{18,19,20},
K. Nandra\inst{1}
}
\titlerunning{{\sc{CircleZ}}  photo-z for AGN}
\authorrunning{Saxena et al.}
\institute{Max-Planck-Institut f\"ur extraterrestrische Physik, Giessenbachstr. 1, 85748 Garching, Germany
\and
Technical University of Munich
Department of Computer Science - I26
Boltzmannstr. 3
85748 Garching b. M\"unchen
Germany
\and 
Exzellenzcluster ORIGINS, Boltzmannstr. 2, D-85748 Garching, Germany 
\and
 LMU Munich, Arnold Sommerfeld Center for Theoretical Physics, Theresienstr. 37 80333 M\"unchen Germany
\and
LMU Munich, Universit\"at-Sternwarte, Scheinerstr. 1, 81679 M\"unchen, Germany
\and
Cerro Tololo Inter-American Observatory/NSF’s NOIRLab, Casilla 603, La Serena, Chile
\and
Perimeter Institute for Theoretical Physics, 31 Caroline St. North, Waterloo, ON N2L 2Y5, Canada
\and
Lawrence Berkeley National Laboratory, 1 Cyclotron Road, Berkeley, CA 94720, USA
\and
National Optical Astronomy Observatory, 950 N. Cherry Avenue, Tucson, AZ 85719, USA
\and
NSF National Optical/Infrared Research Laboratory, 950 N. Cherry Ave, Tucson, AZ 85719, USA
\and
Department of Physics \& Astronomy, University of Wyoming, 1000 E. University, Department 3905, Laramie, WY 8207, USA
\and
Instituto de Estudios Astrof\'isicos, Facultad de Ingenier\'ia y Ciencias, Universidad Diego Portales, Av. Ej\'ercito Libertador 441, Santiago, Chile.
\and
Instituto de Estudios Astrof\'isicos, Universidad Diego Portales, Av. Ej\'ercito Libertador 441, Santiago 8370191, Chile
\and
Kavli Institute for Astronomy and Astrophysics, Peking University, Beijing 100871, China
\and
N\'ucleo de Astronomía de la Facultad de Ingeniería, Universidad Diego Portales, Av. Ej\'ercito Libertador 441, Santiago 22, Chile
\and
Space Telescope Science Institute, 3700 San Martin Drive, Baltimore, MD 21218, USA
\and
Department of Astronomy, University of Washington, Box 351580, Seattle, WA, 98195, USA
\and
Department of Astronomy, University of Illinois at Urbana-Champaign, Urbana, IL 61801, USA
\and
National Center for Supercomputing Applications, University of Illinois at Urbana-Champaign, Urbana, IL 61801, USA
\and
Center for Artificial Intelligence Innovation, University of Illinois at Urbana-Champaign, 1205 West Clark Street, Urbana, IL 61801, USA 
}
\date{Received March XX, 2020; accepted XX, 2020}


  \abstract
   {Photometric redshifts for AGN (galaxies hosting an accreting supermassive black hole in their center) are notoriously challenging 
   and currently better computed via SED fitting,
   assuming that deep photometry for many wavelengths 
   is available. 
   However, for AGN detected all-sky, the photometry is limited and provided by different projects.
   This makes the task of homogenising the data challenging and is a dramatic drawback for the millions of AGN that wide surveys like SRG/eROSITA will detect.}
   {This work aims to compute reliable photometric redshifts for X-ray-detected AGN using only one dataset that covers a large area: the 10th Data Release of the Imaging Legacy Survey (LS10) for DESI. LS10 provides deep \textit{grizW1-W4} forced photometry within various apertures over the footprint of the eROSITA-DE survey. This avoids issues related to the cross-calibration of surveys.
   }
   {We present the results from {\sc{CircleZ}}, a machine-learning algorithm based on a Fully Connected Neural Network.
   {\sc{CircleZ}}  is built on a training sample of 14,000 X-ray-detected AGN  and utilises multi-aperture photometry, mapping the light distribution of the sources.
   }
   {The accuracy ($\sigma_{\rm NMAD}$)  and the fraction of outliers ($\eta$) reached in a test sample of 2913 AGN is 0.067 and 11.6\%, respectively. The results are comparable to or better than those obtained previously for the same field but with much less effort.
   We further tested the stability of the results by computing the photometric redshifts for the sources detected in CSC2 and  {\it Chandra}-COSMOS Legacy, reaching comparable accuracy as in eFEDS when limiting the magnitude of the counterparts to the depth of LS10.}
{The method applies to fainter samples of AGN using deeper optical data from future surveys (e.g., LSST, {\it Euclid}), granted LS10-like information on the light distribution beyond a morphological type is provided.
 With the paper, we release an updated version of the photometric redshifts (including errors and probability distribution function) for eROSITA/eFEDS.}
  
   \keywords{Galaxies: active --
   Galaxies: evolution --
                Surveys --
                Methods: statistical--
                Techniques: photometric
               }

\maketitle
%

\section{Introduction \label{sec:intro}}

It is now accepted that in their lifetime, most (of the massive) galaxies will go through phase(s) during which the supermassive black hole (SMBH) that they host is accreting at a significant fraction of its Eddington limit \citep[e.g.,][]{MadauDickinson2014}. While simulations show that feedback from these Active Galactic Nuclei is necessary to regulate star-formation in their host galaxies \citep[e.g.,][]{DiMatteo2005,Croton2006}, observationally, the interplay between the host and the SMBH  is, however, still a matter of study. Further progress requires that diverse samples of AGNs are studied in detail, so accurate redshift estimates are necessary.
While the number of multi-slit/multifiber spectrographs mounted on large telescopes is ever increasing (e.g., SDSS: Kollmeier et al., in prep; DESI:\citet{Levi2013}; PFS:\citet{Takada2014}; Moons:\citet{Cirasuolo2011}; 4MOST: \citet{deJong2019}) thus raising the completeness of the spectroscopic follow-up of AGN, the urgent need to develop algorithms for the computation of reliable photometric redshifts (photo-z) for these type of objects remains; deep photometric surveys like {\it Euclid} \citep[][]{Mellier2024} and LSST \citep{Ivezic2019} will provide huge samples of AGN at a speed and depth that will not be matched by the new spectrographs!. Photo-zs for AGN  are challenging because the relative host/nucleus contribution to the emission is unknown a priori and even a modest contribution to the host galaxy SEDs by the continuum emission from the AGN hides the features used for identifying the redshift. The net result is a large fraction of sources having multiple possible redshifts (degeneracies).

Both machine learning and SED fitting have been adopted \citep[see][for a review]{Salvato18b} in an attempt to improve on the results (e.g., Table \ref{tab:history}). What we have learned is that a) a large number of photometric points is needed to break such degeneracies \citep[e.g.,][]{Brescia2019}; b) photometry from narrow and intermediate bands is more valuable as opposed to mid-infrared broadband photometry given their increased ability in pinpointing the emission lines and their intensity, separating thus starforming from AGN powered objects. \citep[][]{Salvato09,Cardamone10,Luo10,Hsu14}; c) machine learning can reach high accuracies for sources represented by the training sample at the cost of a very high fraction of outliers for sources outside the parameter space \citep[e.g.,][]{Duncan18, Norris2019, Salvato22}; d) SED fitting performs well, granted that a redshift prior (expressed for example via a redshift-dependent absolute magnitude range) is applied and the templates are appropriately selected \citep{Luo10, Fotopoulou12, Hsu14, Ananna2017} e) due to the presence of more features in the SED (e.g., Lyman break, Balmer break), the quality of the photo-zs for inactive galaxies (computed either via SED fitting or via machine learning) remains superior to the photo-zs for AGN; f) for AGN dominated sources ( i.e, X-ray bright) reliable redshifts are more difficult to be obtained \citep{Brescia2019}.

All the works mentioned above relied on deep, PSF-homogenised photometry in many bands across the electromagnetic spectrum in pencil-beam areas, which is challenging to achieve for wide- or all-sky area surveys.


\begin{table*}
\centering 
\caption{Accuracy and fraction of outliers obtained for X-ray surveys at different X-ray depths}
\small
\begin{tabular}{lcccccc}
\hline
\hline
 Survey& Flux limit & Area & Photometry & Accuracy & Fraction of outliers & Method \\
Ref.    &${\rm erg/s/cm^2 }$ & ${\rm deg^2}$ &           &$\sigma_{\rm NMAD}$ & $\eta$ & \\
\hline
\hline
Chandra-COSMOS  & $1.9 \times 10^{-16}$ &0.9 & N/I/B, Uv-O-NIR-MIR &0.015 & 6\% & SED fitting \\
\citep{Salvato09} & & & && &\\
\hline
XMM-COSMOS  & $1.17 \times 10^{-15}$ & 2.13 & N/I/B, Uv-O-NIR-MIR &0.015 & 6.3\% & SED fitting \\
\citep {Salvato11} &&&&&& \\
\hline
CDFS  &$9.1 \times 10^{-18}$ & 0.13 & N/I/B, Uv-O-NIR-MIR &0.014 & 6.7\% & SED fitting \\
\citep{Hsu14, Luo10} &&&&&&\\
\hline
Stripe 82X XMM archival  &$7 \times 10^{-17}$ & 7.4 &B, UV-O-NIR-MIR& 0.06 & 13.2\% & SED fitting \\
\citep{Ananna2017} &&&&&& \\
\hline
Stripe 82X (XMM archival)&$7 \times 10^{-17}$ & 7.4 &B, UV-O-NIR-MIR & 0.06 & 12.7\% & ML \\
\citep{Brescia2019} &&&&&& \\
\hline
Stripe 82X (XMM archival) & $1 \times 10^{-14}$ &7.4 &B, UV-O-NIR-MIR& 0.122 & 13.1\% & ML \\
\citep{Brescia2019} &&&&&& \\
\hline
XMM-SERVS & $1 \times 10^{-14}$&4.5 & B, Uv-O-NIR-MIR&0.04&9.2\% & SED fitting \\
\citep{Ni2021} &&&&&&\\
\hline
XMM-SERVS & $1.3 \times 10^{-14}$ &3.0& B, Uv-O-NIR-MIR&0.04&9.6\% & SED fitting \\
\citep{Ni2021} &&&&&& \\
\hline
\hline
\end{tabular}
\tablefoot{Modified version of the table presented in \citet{Ananna2017}. The X-ray flux is measured in the 0.5-2 keV range (0.5-10 keV for XMM-SERVS). In addition, we report the area coverage, type of photometry available (Narrow, Intermediate, Broad band photometry), and  wavelength coverage (UV, Optical, Near-Infrared, Mid-Infrared).}
 \label{tab:history}      
 \end{table*}
In fact, in the computation of photo-zs for the about 22000 AGN detected by eROSITA \citep{Predehl2021, Merloni2024} in eFEDS  \citet{Salvato22} already pointed out how the inhomogeneity of the data quality drastically affects the reliability of the results\footnote{See section 7.2 of \citep{Salvato22}, and Table \ref{tab:accuracyBreackdown} of this paper, where the quality of the photo-zs in the area within and outside the homogenized optical-NIR photometry from KiDS \citep{Kuijken19} is compared}. Using the traditional method, for eRASS:8 in the best case scenario, we expect results not better than an accuracy of 0.07 and a fraction of outliers of 25\%, after an extenuating effort of homogenization of the photometry from various surveys covering from UV to Mid-infrared.
Does this showcase significant issues for computing photo-zs for AGN detected by the eROSITA all-sky surveys?

Certainly, it would if we tried to determine the redshift using the total fluxes emitted by the AGN (and errors) as the only input, as it is done standardly.\\
We have already mentioned that the extension of the sources in optical images provides a prior on the possible redshift solutions \citep[][]{Salvato09}. We can go beyond the differentiation between extended and point-like sources using Machine Learning.
The feature analysis of all parameters available in the SDSS DR9 \citep{Ahn2012} and SDSS DR7 \citep[][]{Abazajian09} of spectroscopically confirmed AGN and galaxies presented in \citep{D'Isanto18}, indicated that the photo-zs quality could have been improved (e.g., smaller bias, smaller root mean square error; RMSE) for the photo-zs computed using fluxes and colors, by considering features such as Petrosian magnitudes\footnote{Flux measured within a circular aperture whose radius is defined by the shape of the azimuthally averaged light profile}. In particular, what was found is that ratios between magnitudes computed assuming different models (point source, exponential, de Vaoucouler, or the combination of exponential and de Vaucouler) carry relevant information.
Similarly, in \citet{Nischizawa2020}, the computation of photo-zs for extragalactic sources detected in the HSC survey \citep[][]{Aihara18a}  was carried out with Machine Learning relying solely on the optical photometry measured at 1.3 and 1.5 arcsecond apertures, computed assuming a PSF, an exponential, and a de Vaucouler profile. The accuracy reached by this approach (although limited to the parameter space covered by their training sample) was comparable to or even better than the results obtained via SED fitting after combining optical data with UV, near-infrared, and mid-infrared data \citep[see][for a direct comparison]{Salvato22}.  
Recent works are also trying to compute the photo-zs using the images directly \cite[e.g.,][Roster et al., in prep]{Shuldt2021}.
All these facts indicate value when considering the morphology, the size, and, in particular, the light distribution (e.g., in the form of an azimuth profile) when computing the photo-zs. This is because all these quantities are, at least partially, redshift-dependent.
In particular, unlike the single total {\it model} flux per each photometric band (which is what most photometric surveys provide), the fluxes within different apertures and/or assuming different model profiles of light distribution (exponential, Sersic, Point-like, etc.) carry with them information on the host/nucleus relative contribution.

This is confirmed by the work presented in \citet{Zhou2021,Zhou2023}. There, photo-zs for DESI sources using the data release 9  of the Legacy Imaging Survey and considering the magnitudes, the colors g-r, r-z, z-W1, half-light radius, axial ratio (ratio between semi-minor and semi-major axes), and a model weight, were computed. While the accuracy was extremely good for galaxies (below 0.02 in accuracy and fraction of outliers below 10\%, depending on the sample), the accuracy reached for AGN and QSO is poor, with an accuracy of 0.07 and up to 33\% outliers ($\eta$\footnote{$\eta =\lvert z_{\rm phot}-z_{\rm spec} \rvert>0.15 \cdot (1+z_{\rm spec})$}) and a low accuracy ($\sigma_{\rm NMAD}$\footnote{$\sigma_{\rm NMAD}=1.48 \cdot median((z_{\rm phot}-z_{\rm spec})/(1+z_{\rm spec}))$}) of 0.08\footnote{updated results from the same authors using LS10 are described at \url{https://www.legacysurvey.org/dr10/files/\#photo-z-sweeps-10-1-photo-z-sweep-brickmin-brickmax-pz-fits}}. 

The latest Legacy Survey imaging for DESI Data Release \citep[LS10][]{Dey2019},  available since January 2023, and a newly available large sample of spectroscopically confirmed AGN (see Section \ref{SEC:training}) allowed us to test this hypothesis further.
Here, we present the results from {\sc{CircleZ}}, a machine learning algorithm based on a fully-connected neural network (FCCN) that reliably computes photo-zs for AGN using only the data available from Legacy Survey without adding additional information from external surveys at different wavelengths. More than in the algorithm \citep[very similar to {\sc{DNNZ}};][]{Nischizawa2020}, the innovation here is on the use of new features (colors within apertures and the quality of the fit to the azimuth profile assuming different galaxy profile models) that, just as total fluxes, themselves correlate with redshift. The increase in useful features allows us to break the degeneracy in the redshift solution typical of photo-zs for AGN computed with a limited number of photometric bands. We demonstrate this by recomputing the photo-zs of eROSITA/eFEDS and comparing the results previously obtained via SED fitting with {\sc{LePHARE}}\citep{Arnouts99, Ilbert06, Salvato09} and via machine learning with {\sc{DNNZ}} \citep{Salvato22}. In addition, we tested the quality of the photo-zs for AGN selected in the Chandra Source Catalog Release 2 (CSC2\footnote{\url{https://cxc.cfa.harvard.edu/csc/}} and Legacy {\it Chandra}-COSMOS (a deep pencil-beam survey) for those sources with secure spectroscopic redshift. We show that for AGN with the counterparts of brightness typical of LS10, the photo-zs are reliable, suggesting that the method could also be adopted for fainter sources, granted that features and training samples are available. This is relevant for surveys such as {\it Euclid} \citep[][]{Mellier2024} and LSST \citep{Ivezic19}.

In Section \ref{sec:Data}, we present the ancillary data and the features that will be adopted,  while in Section \ref{SEC:training}, we describe the training sample. Section \ref{section:ARCH} presents the architecture of the algorithm and can be skipped by the reader interested only in the results of the application.
In Sec \ref{sec:eFEDSquality} we recompute the photo-zs for the extragalactic sources detected by eROSITA \citep{Predehl2021, Merloni2024} in the eFEDS field \citep[][]{Brunner2022, Salvato22} so that a direct comparison with the results from {\sc{DNNZ}} \citep[from][]{Nischizawa2020} and {\sc{LePHARE}} \citep[from][]{Salvato22} can be made. 
Finally, in Section \ref{sec:comparison}, the ability of {\sc{CircleZ}}  in computing photo-zs for generic X-ray surveys is tested using the AGN detected in CSC2 and {\it Chandra}-COSMOS Legacy \citep[][]{Civano12, Marchesi16}. The discussion is presented in Section~\ref{sec:discussion}, while the summary and a few considerations about the new photometric surveys available to the community and photo-zs' future will conclude the paper with Section \ref{sec:Conclusions}.
For the interested reader, further analysis on the importance of the features used by {\sc{CircleZ}} for computing photo-zs is presented in Appendix~\ref{sec:FeaturesRelevance}. 
\section{The Photometric Data
\label{sec:Data}}
Typically, photometric observations are confined to survey-specific patches on-sky, each equipped with its unique configuration of broad and narrow band filters and varying transmission profiles.
For samples of sources defined over areas of the sky larger than the typical photometric survey-specific patches, the inhomogeneity in observational setups can introduce inconsistencies when attempting to standardize surveys and establish a unified framework that enables seamless integration and comparison across different survey datasets. The already unstable photo-zs computed for AGN suffer from this inhomogeneity more than the photo-zs computed for galaxies, so all-sky surveys of AGN, like the samples generated from the eROSITA catalogs, would not have a homogeneous quality.

\begin{figure*}
\centering
\includegraphics[width=9.1cm]{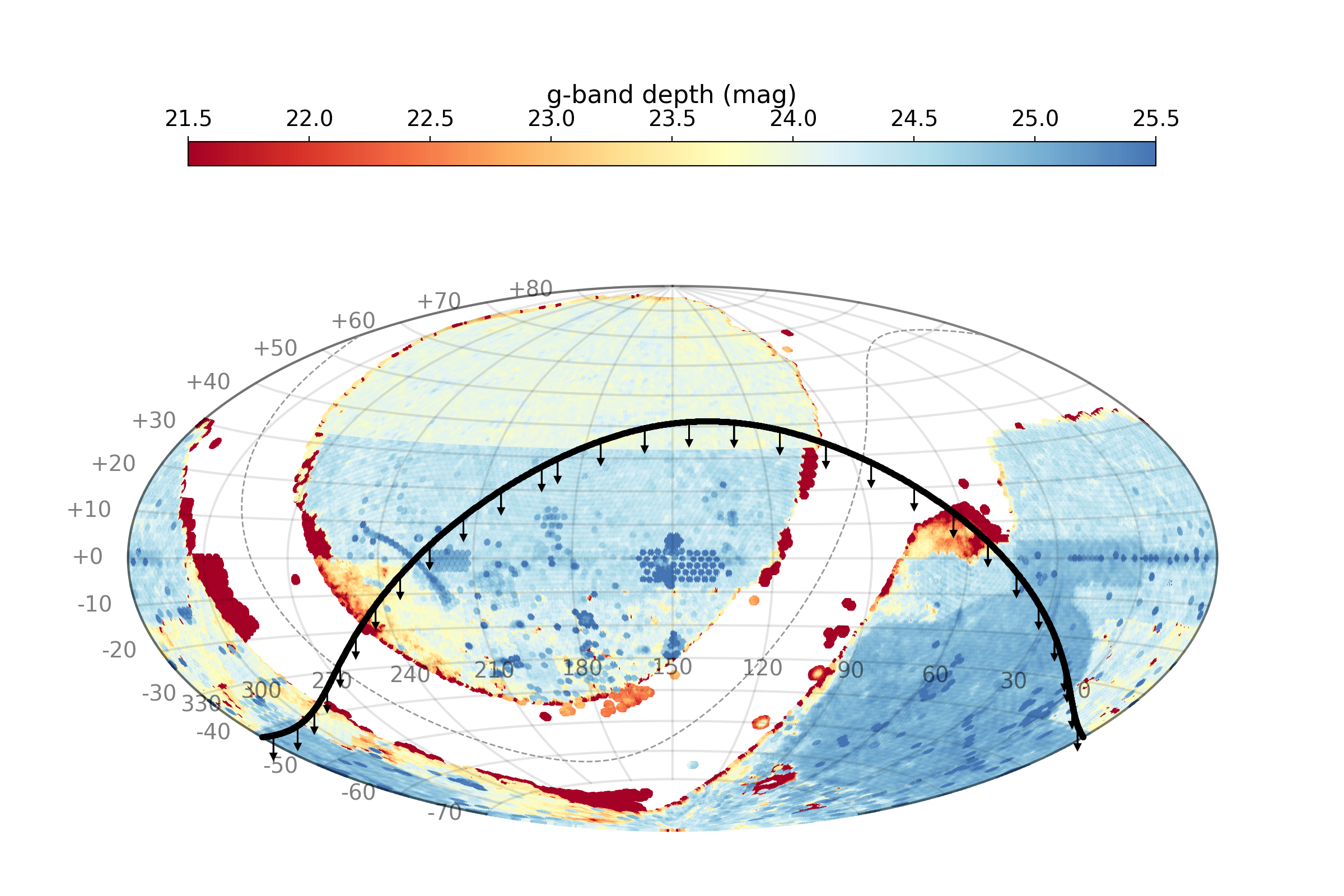}
\includegraphics[width=9.1cm]{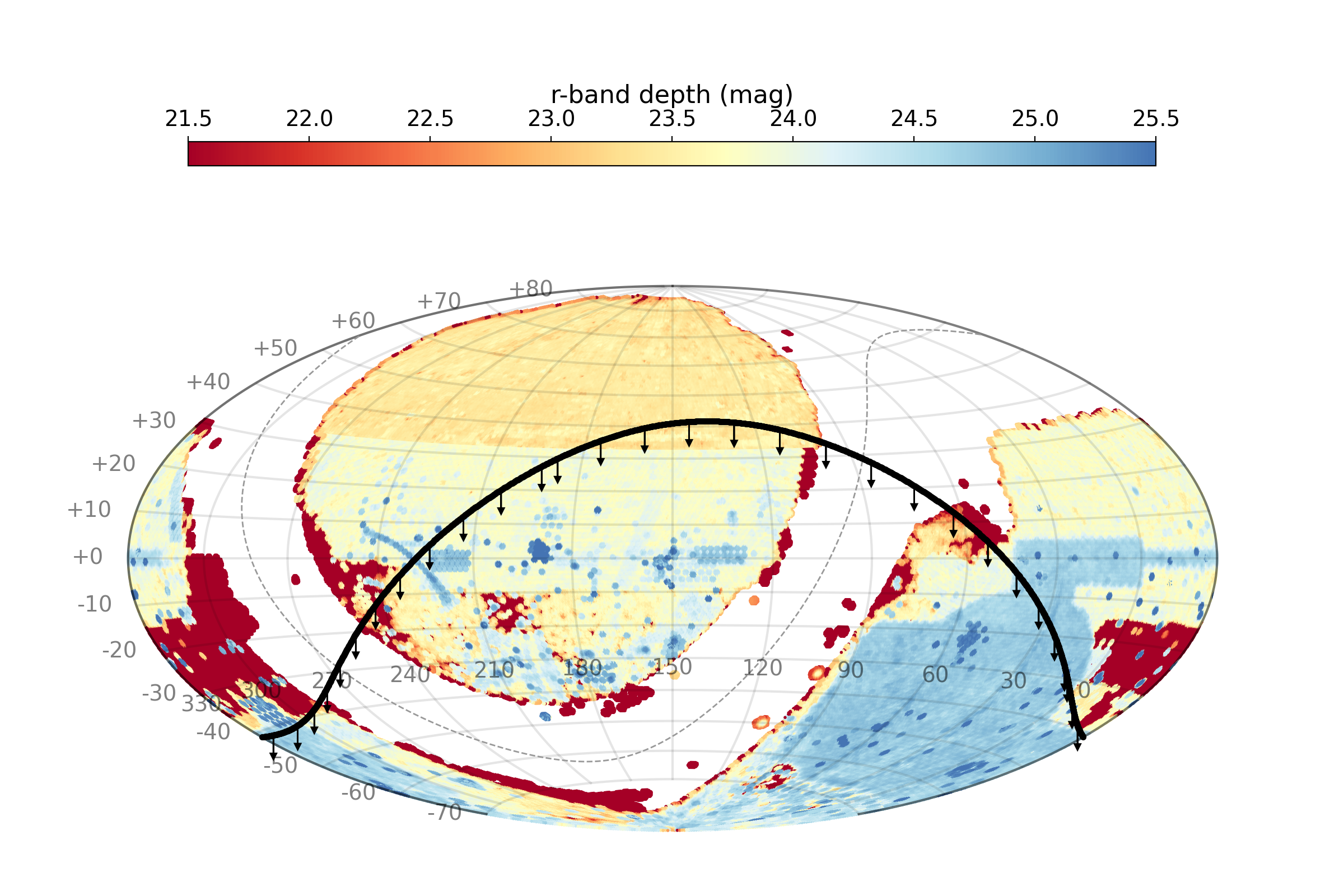}
\includegraphics[width=9.1cm]{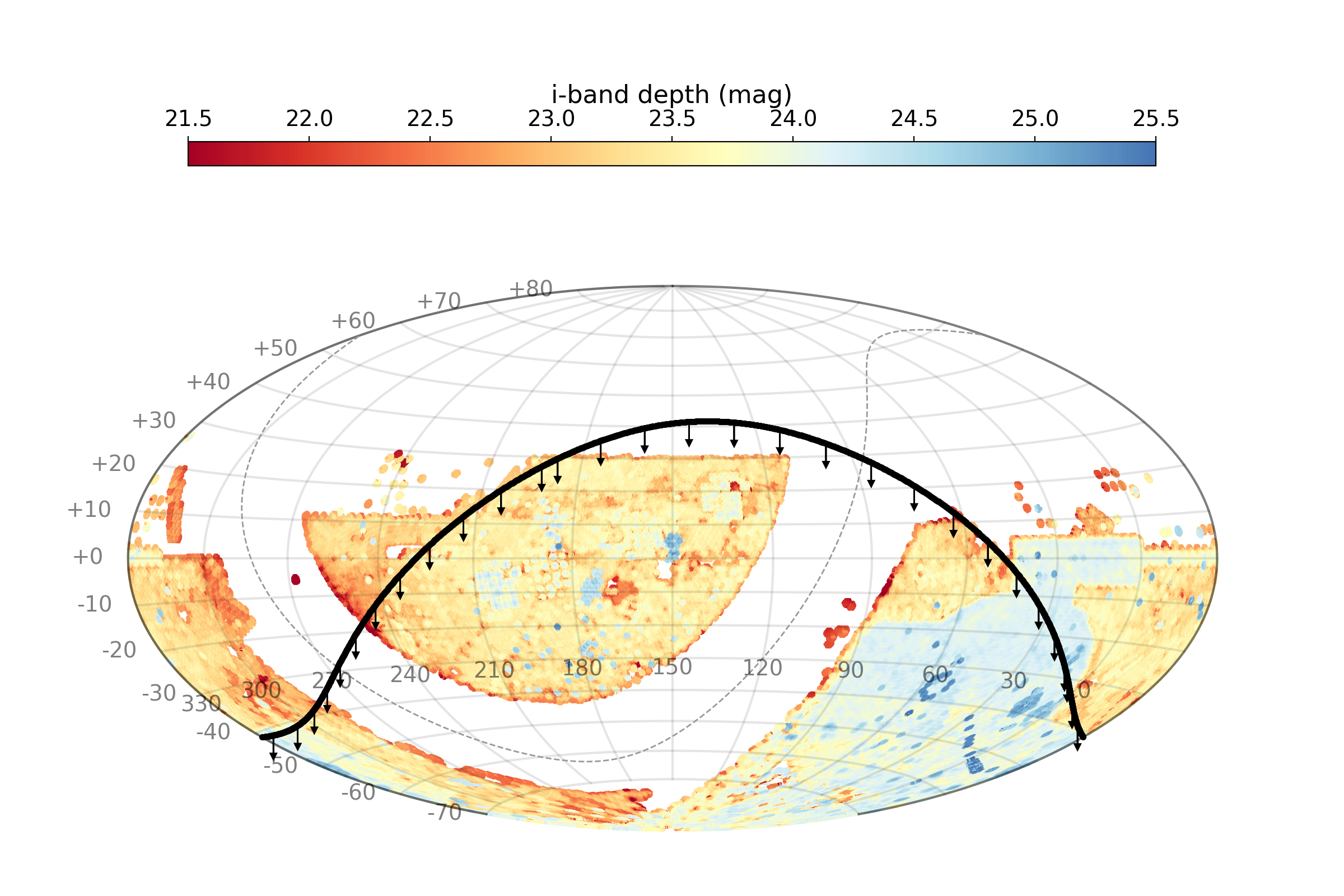}
\includegraphics[width=9.1cm]{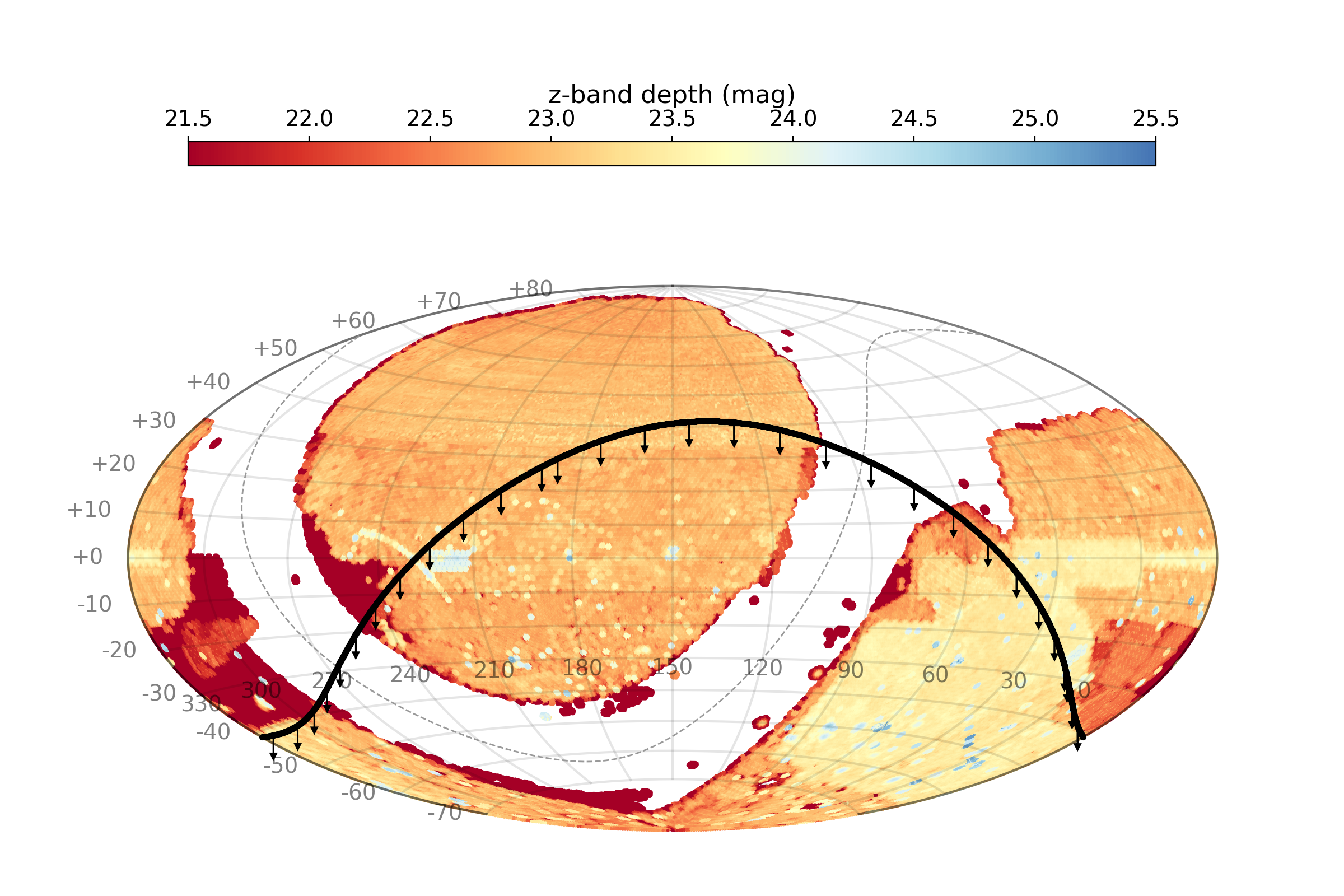}
\caption{Depth and coverage of LS10 in, clockwise, {\it g,r,i,z} filters. The black line delimits the eROSITA-DE area. The use of a common scale highlights the depth inhomogeneity of the data. For the rest of the paper, we use LS10-South (declination < 32.375$\deg$), which is the area covered by the {\it i} band and also within the footprint of eROSITA.}
\label{fig:LS10Depth}
\end{figure*}

Fortunately, the Legacy imaging Survey for DESI \citep[][]{Dey2019} data release 10 (LS10) provides registered, background-subtracted, and calibrated "all-sky" photometry, covering almost entirely the eROSITA-DE footprint. Despite the coverage being limited to 4 optical bands (see Fig \ref{fig:LS10Depth}) and WISE, it provides photometry that goes beyond the total flux measured assuming a galaxy profile, and the question is whether access to these features is sufficient to estimate reliable photo-zs for AGN, without the need of combining the data with Galex in UV \citep{Bianchi14} and VHS\citep{McMahon2013} or UKIDSS\citep{Laurence2007} in NIR, thus simplifying the work dramatically.  
Interested readers can refer to the "Content" section of LS10\footnote{\url{https://www.legacysurvey.org/dr10/description/}} for comprehensive information on the survey data, processing details, and coverage. In the following sections, we explore the features considered for {\sc{CircleZ}}.

\subsection{Going beyond total, model fluxes \label{subsec:features}}

In this paper, we demonstrate that reliable photo-zs for AGN with a limited number of photometric points are achievable by using a) machine learning, b) a training sample limited to AGN and, more specifically, to X-ray detected AGN, c) considering parameters that correlate with redshift,  in addition to standard parameters (magnitudes, colors, and morphology\footnote{Meant here as an extension in images and classification}). We identify the parameters as those that fully describe the light distribution of a source in its components (nuclear, bulge, disk) and are already available in the LS10 catalog. We describe the construction of the training sample in the next section, while here, we focus on describing the parameters (features) considered.

LS10 is composed of two distinct parts, LS10-North and LS10-South. LS10-north is obtained by combining two optical ({\it g,r,z}) surveys (Beijing-Arizona Sky Survey--BASS\footnote{\url{https://www.legacysurvey.org/bass/}}; Mayall {\it z}-band Legacy Survey-- MzLS\footnote{\url{https://www.legacysurvey.org/mzls/}} in addition to the data obtained for DESI\footnote{\url{https://www.desi.lbl.gov/the-desi-survey/}} with the Dark Energy Camera (DECam) at the CTIO Blanco 4-m telescope. It corresponds to the northern portion of the Data Release 9 \citep{Dey2019}.  LS10-South, compared to the Data Release 9,  has deeper data taken with DECam in {\it g,r,i,z}. For the rest of the paper, we will refer to LS10-South as LS10, as the area covered by the {\it i} band (see Figure \ref{fig:LS10Depth}), which almost completely maps the eROSITA-DE area. In particular, the data include all the observations taken within the DeROSITAs program (P.I. A. Zenteno; Zenteno et al., in prep).
For each source detected in the four optical bands, LS10\footnote{The complete list of parameters and their description are available at {\url https://www.legacysurvey.org/dr10/catalogs/}} provides PSF-forced photometry (and related errors) in the optical {\it g,r,i,z}  and mid-infrared (W1, W2, W3 and W4) from all imaging through year 7 of NEOWISE-Reactivation, computed in multiple apertures (0.5, 0.75, 1.0, 1.5, 2.0, 3.5, 5.0, 7.0 arcsecond radii in the optical and 3, 5, 7, 9, 11 arcseconds radii in the mid-infrared). It also provides the total flux for each band, assuming the best model (indicated in the column TYPE: PSF"=stellar, "REX"="round exponential galaxy", "DEV"=deVauc, "EXP"=exponential, "SER"=Sersic profile) fitted to the azimuth profile in the {\it r} band, the quality of the fit for each model, together with an extensive catalog of observational attributes and flags indicating problematic observations.

\begin{figure*}
\centering
\includegraphics[width=5cm]{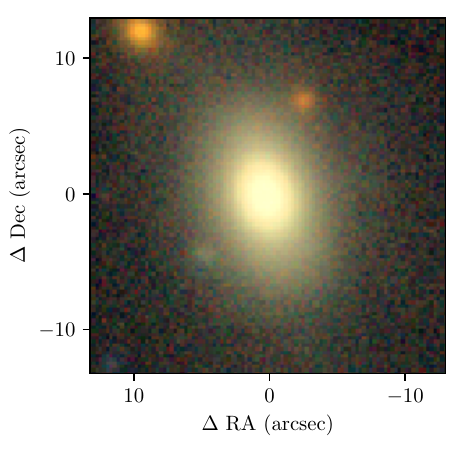}
\includegraphics[width=5cm]{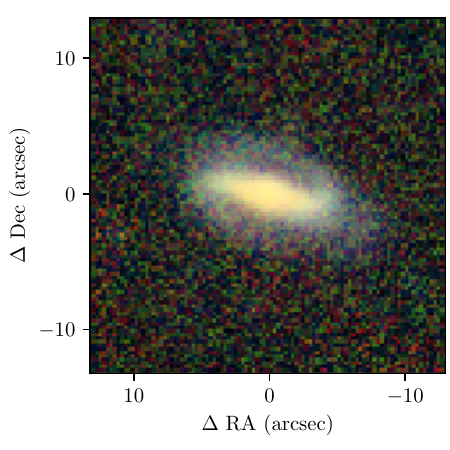}
\includegraphics[width=5cm]{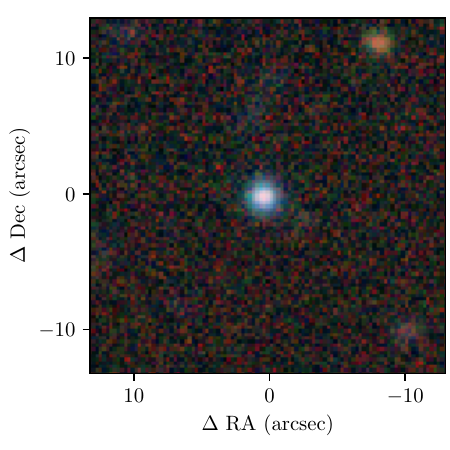}
\includegraphics[width=5cm]{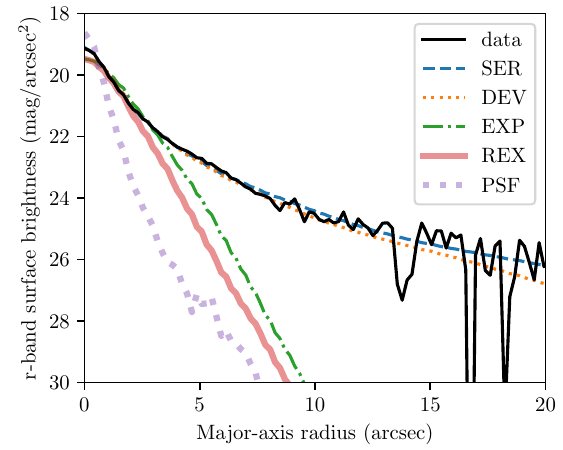}
\includegraphics[width=5cm]{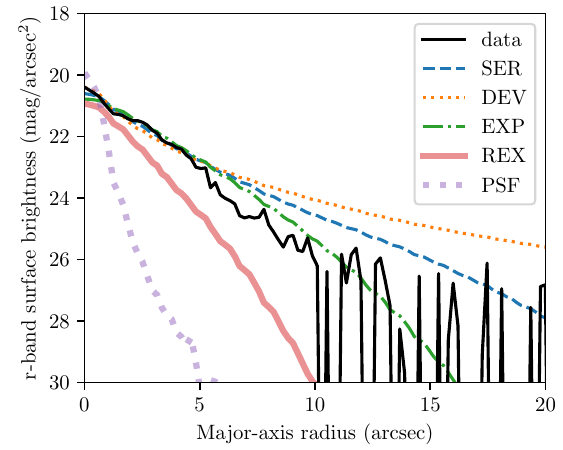}
\includegraphics[width=5cm]{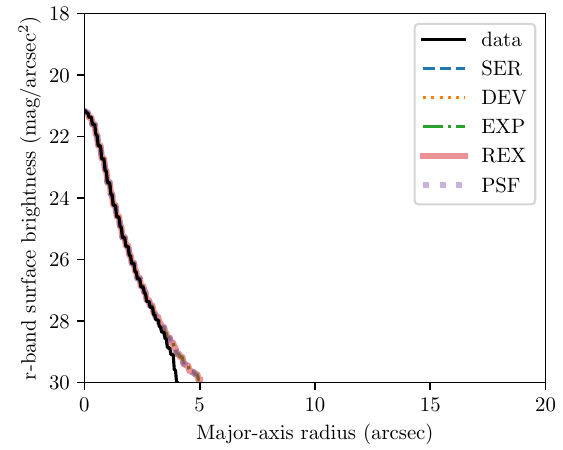}
\includegraphics[width=5cm]{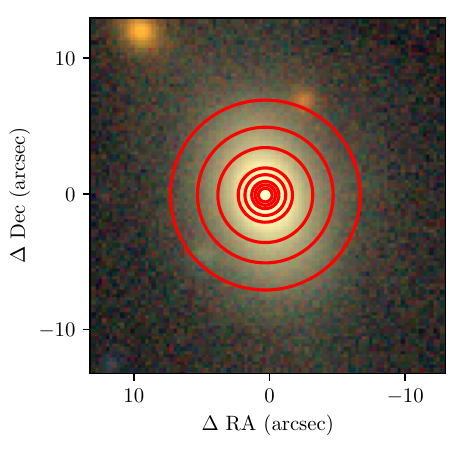}
\includegraphics[width=5cm]{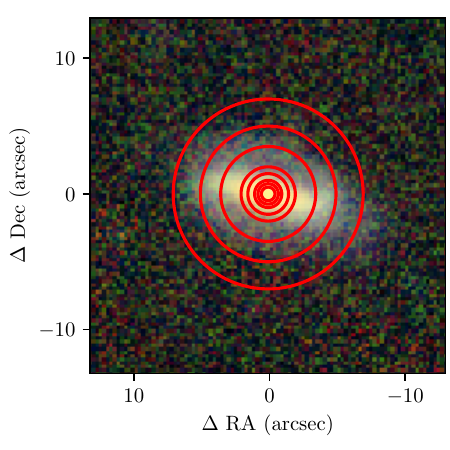}
\includegraphics[width=5cm]{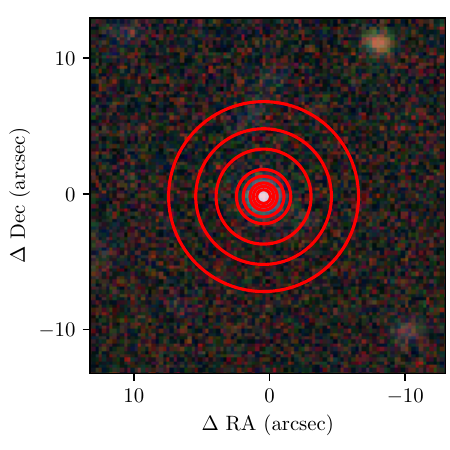}
\includegraphics[width=5cm]{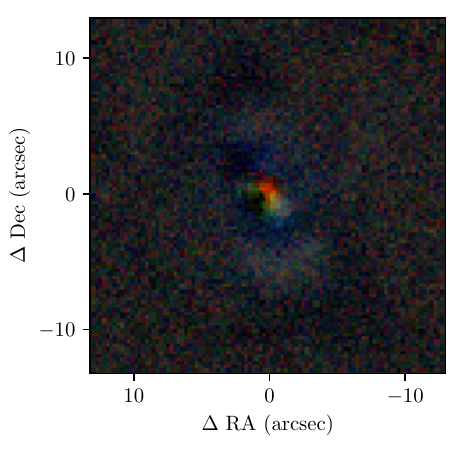}
\includegraphics[width=5cm]{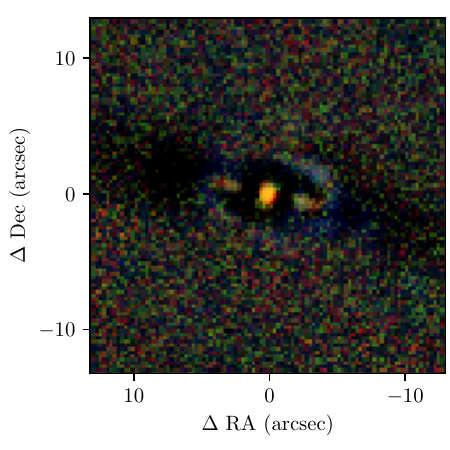}
\includegraphics[width=5cm]{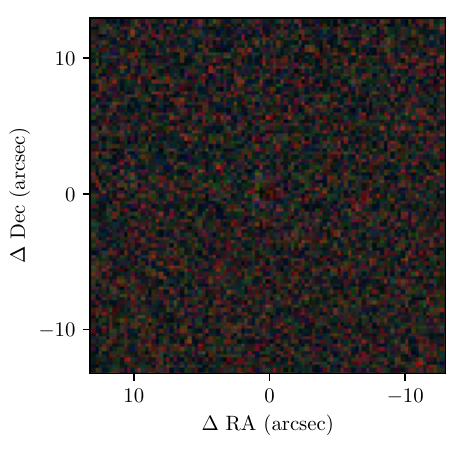}
\includegraphics[width=5cm]{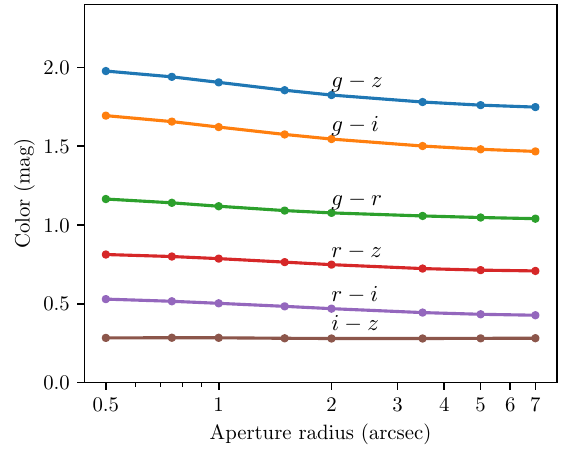}
\includegraphics[width=5cm]{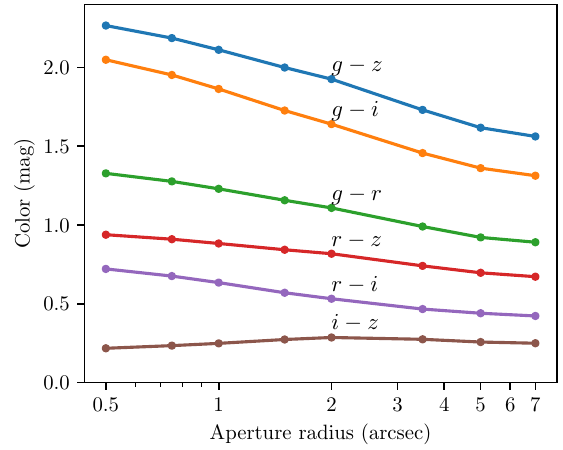}
\includegraphics[width=5cm]{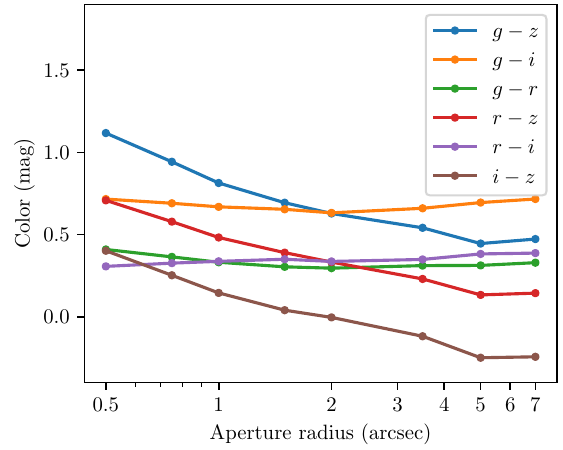}
\caption{Top: Optical RGB image of an inactive galaxy (Left) and a low-redshift X-ray detected AGN (Middle) and a QSO (Right). Second row: azimuth profile of the sources, with over-plotted available model profiles (PSF, REX, EXP, DEV, and SER). Third row: the apertures within which ap\_flux[band]  is measured.
Fourth row: the residual image after subtracting the best model. Last row: within each aperture, all colors are computed; the figures show how the profiles of the colors change with the distance from the center, depending on the source type (see more details in the main text.}
\label{fig:LS10Features}
\end{figure*}

Figure \ref{fig:LS10Features} provides a visual representation of the features used in our analysis and demonstrates the method's strength. The figure shows the RGB image of a normal galaxy, a local AGN, and a QSO. In the Second row, the azimuth profile of the sources is shown, with over-plotted available model profiles (PSF, REX, EXP, DEV, and SER). The quality of the fits is encapsulated in the column "dchisq" of the catalog, while the best-fitting model is used to classify the source, specified in the column "TYPE". In these examples, the galaxy on the left can be equally well-fitted by the "DEV" and "SER" profiles, with "SER" being slightly better. For the AGN, in the middle panel, none of the models is a good profile representation, with "EXP" being selected. This bad fitting would reflect in the quality of the photo-zs computed with algorithms that use model\_flux photometry only. For each source, the flux and its residual, after removing the best model, is measured within the apertures shown in the third row.  Within each aperture, all colors are computed, and the last row shows how the profiles of the colors change with the distance from the center, depending on the source type. This dependence on the size is implicitly used to compute the photo-zs correctly. For the interested reader, the importance of these features and their correlation with the redshifts are further discussed in the Appendix \ref{sec:FeaturesRelevance}.

In summary, the features extracted from LS10 and considered in this work are:\\
FLUX\_[band],  FLUX\_IVAR\_[band], FLUX\_res\_[band]: these are the fluxes, inverse variance, and residuals, measured assuming the best  model profile;\\
APFLUX\_[band],~APFLUX\_IVAR\_[band],~APFLUX\_res\_[band]: these are the same as above but this time measured on the images, within various apertures;\\  
TYPE: the best model of the light profile. Originally this is a string that we converted to numeric.\\
dchisq\footnote{more details are available at \url{https://www.legacysurvey.org/dr10/catalogs/}}: Difference in $\chi^2$  between successively more-complex model fits: PSF, REX, DEV, EXP, SER. The difference is versus no source.

We then pre-processed the data by correcting the total and aperture fluxes for Galactic extinction\footnote{We follow the recipe provided by the LS10 team: Flux\_[band]/MW\_transmission\_[band], with MW\_transmission\_[band] also available in the LS10 catalog}, and computed all possible colors, including colors within each aperture. 
We also computed the signal-to-noise as:\\
$ {\rm SNR\_[band]} = {\rm FLUX\_[band]} \times \sqrt{\rm (FLUX\_IVAR\_[band])}$

\section{Training and blind samples
\label{SEC:training}}

So far, unlike photo-zs computed via SED fitting, supervised machine learning algorithms, such as regression or classification models, which establish mappings from inputs to outputs, have shown to be limited in their ability to extrapolate beyond the physical parameter space covered by the data they were trained on \citep[see more details in][]{Salvato18b}. Thus, their predictive accuracy may diminish significantly when applied to input values that lie well beyond patterns and ranges encountered during training and, therefore, rely extensively on the availability of a representative training sample \citep[][]{Newman2022}. Various populations, such as star-forming galaxies, quiescent galaxies, or AGN, exhibit unique physical attributes, highlighting the need for a thoughtful selection of tailored sources to be included in the training sample \citep[][]{Duncan18}. Given our objective to comprehensively explore the whole plethora of eROSITA-detected extragalactic populations, we have constructed a training sample that extensively covers the diversity in the AGN population.
The training sample spans from low-luminosity, obscured, type II AGN \citep[][]{Kauffmann2003, Laor2008} to type I broad-line AGN and quasars \citep[][]{Reynolds1997, Richards2006} encompassing a wide redshift range, up to z=6.6.

\subsection{Construction of the sample for training
\label{agn_cats}}
We are particularly interested in computing the photo-zs for X-ray-detected AGN, particularly those detected by eROSITA. By nature of being all-sky, the eROSITA survey is shallower than the pencil-beam surveys conducted with XMM and {\it Chandra}, resulting in a sample less representative of galaxy-dominated and obscured AGN and richer in bright and nearby, resolved AGN. For this reason, we created the training sample by combining sources from eROSITA/eFEDS 
\citep[][]{Salvato22} and 4XMM \citep{Webb2020} for which reliable multi-wavelength counterparts from LS10 have been identified following the same procedure described in Section 3.1 and Appendix A of \citep[][]{Salvato22} for 3XMM. \\
The counterparts to the training sample are then matched in Right Ascension and Declination (max separation of 1 arcsecond) with a compilation of spectroscopic redshifts available within the eROSITA collaboration.
At this stage, the compilation is a mere list of catalogs from the literature collated. It encompasses approximately 13 million celestial sources, thus including duplicates and contradicting redshift measurements \citep[see][for more details]{Kluge2024}. 
Additionally, we extend the compilation by incorporating QSOs with reliable redshifts from the low-resolution spectra in Gaia DR3 \citep[flagQSOC =0][]{refId0}.
For some sources, multiple spectroscopic matches are available, with discordant results. This can occur, for example, due to wrong line identification in low signal-to-noise spectra. Here, we prioritize redshifts from eFEDS (Aydar et al.,  in prep), given that most have all been visually inspected. Alternatively, a single entry is randomly selected, provided that the difference in spectroscopic redshift between the smallest and largest redshift entry $\Delta z$ does not exceed $\Delta z$ > 0.05. Objects failing to meet this criterion ( $ \leq 0.3\%$ of cases) are excluded from consideration to avoid the accidental assignment of incorrect redshift values to objects in the training sample. Objects with spectroscopic matches falling within the redshift range of $z \leq$ 0.002 are assumed to be galactic and, therefore, excluded from any further consideration. The compilation is enriched by including 152 proprietary entries from SDSS-V (Kollmeier et al., in prep) (SPECPRIMARY == 1, ZWARNING == 0). Within SDSS-V, the Black Hole Mapper (BHM, Anderson et al., in prep) is one of the primary surveys, specializing in long-term, time-domain investigations of AGN and the optical characterization of eROSITA X-rays sources detected during the initial 1.5 years of the all-sky survey. This part of the training sample is represented in the histogram of Figure \ref{fig:redshift} with a solid violet line.

\subsection{Extension of the training sample to high-z sources}
The contribution to the high-redshift AGN population at $z$>5 from X-ray pencil-beam surveys is feeble \footnote{Unsurprisingly, most of the sources at redshift higher than five that we visually inspected were wrong}. By construction, the training sample would not represent the high-$z$ population. However, eROSITA is expected to detect about 90 sources at $z$>5.5 by the end of the mission \citep{Wolf21}. At the same time, about a dozen have already been identified \citep[][]{Medvedev20, Medvedev2021, Wolf21, Wolf2023, Wolf2024}. The optical spectra of these high-z QSOs closely resemble the spectra of QSOs at low redshift \citep[e.g.,][]{Mortlock2011, Yang2021}, which is why we have extended the training sample, adding 283 sources spectroscopically confirmed at  $z>5.5$, from the compilation of \citet{Fan2023}, regardless of whether or not they are detected in the X-rays.
The violet dashed histogram in Figure \ref{fig:redshift}  shows the redshift distribution of these sources. 
\begin{table}
\centering 
\caption{Sources used in this work}
\small
\begin{tabular}{lccc}
\hline
\hline
 Survey & w/ & w/ & Unique \\
 used for training   & LS10 & spec. redshift & \\
 \hline
 \\
4XMM & 32900 & 9881   & 4031 (27.7\%)\\
eFEDS  & 27367& 10309 & 10142 (69.6\%) \\
HIGH-Z AGN & 391 & 391 & 391 (2.7\%)\\
\\
\hline
\hline
Blind samples & & & \\
\hline
\\
CSC2  & 5643& 1654 & 416\\
{\it Chandra}-COSMOS & 2491& 2198 & 1699\\
\\
\hline
\end{tabular}
\tablefoot{The table shows the breakdown distribution of sources used for training (i.e., training, test, and validation samples) and in the blind samples, before and after the match with the spectroscopic redshift compilation, and after cleaning for duplicates. The numbers within the parenthesis show the contribution to the sample used for training.} 

 \label{tab:breakdown}      
 \end{table}
\subsection{Samples for blind tests
\label{subsec:BlindSamples}}

In addition to the sample that will be partitioned in training, test, and validation (see Section \ref{subsec:TTV}), we want to blindly test the consistency of the results obtained by {\sc{CircleZ}} for two additional samples:\\

{\it CSC2}: this blind sample is generated using the counterparts to the CSC2 objects\footnote{\url{https://cxc.cfa.harvard.edu/csc/}} in the same way as described for the training sample. The original sample of CSC2 sources with reliable counterparts included 5643 sources. Screening for sources with spectroscopic redshift and after removing the sources that are in common with the training sample, 413 remain in the area covered by LS10-South, after including 26  redshifts from SDSS-V. \\

{\it Chandra}-COSMOS Legacy: for the counterparts to the X-ray sources presented in \citep[][]{Marchesi16}, we consider the corresponding LS10 source using a match in coordinates (max separation of 1 arcsec). We then matched the sources to the compilation of redshifts available within the COSMOS collaboration (Khostovan et al., in prep), considering only reliable redshifts (Flags 3,4,13,14). After removing the sources common to the training sample, 1699 sources are considered.

\begin{figure}[ht]
\centering
\includegraphics[width=9.5cm]{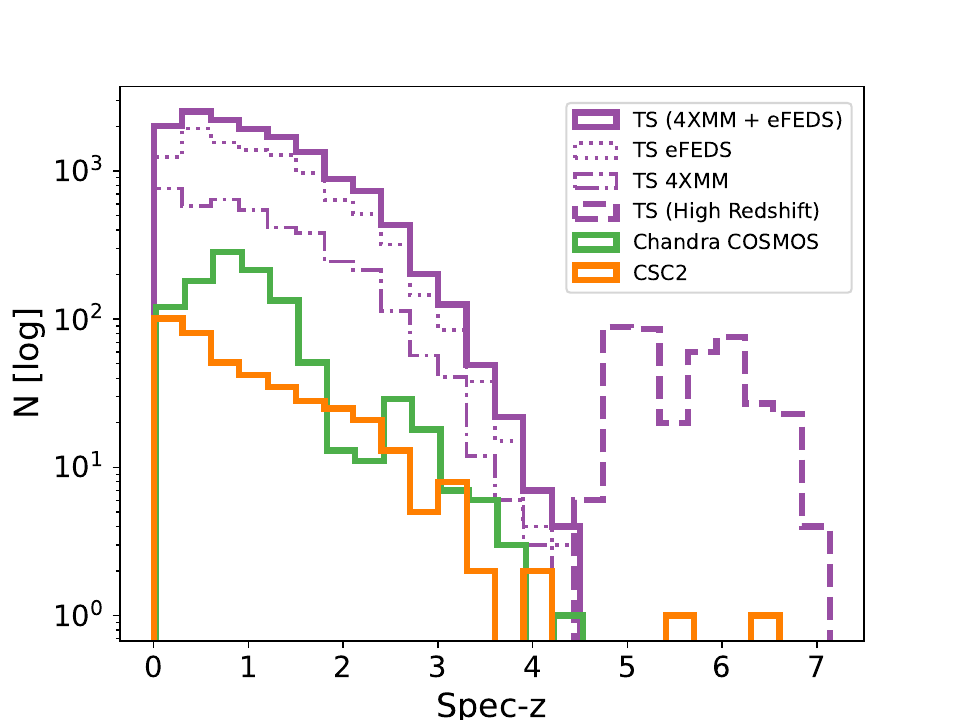} 
\caption{Redshift distribution of the training (violet) and the blind samples (CSC2 in orange; {\it Chandra}-COSMOS Legacy in green). The distribution of the training sample is split between sources from {\it 4XMM}, eROSITa/eFEDS, and high-z sources from \citet{Fan2023}.}
\label{fig:redshift}
\end{figure}

\begin{figure*}[ht]
\centering
\includegraphics[width=6cm]{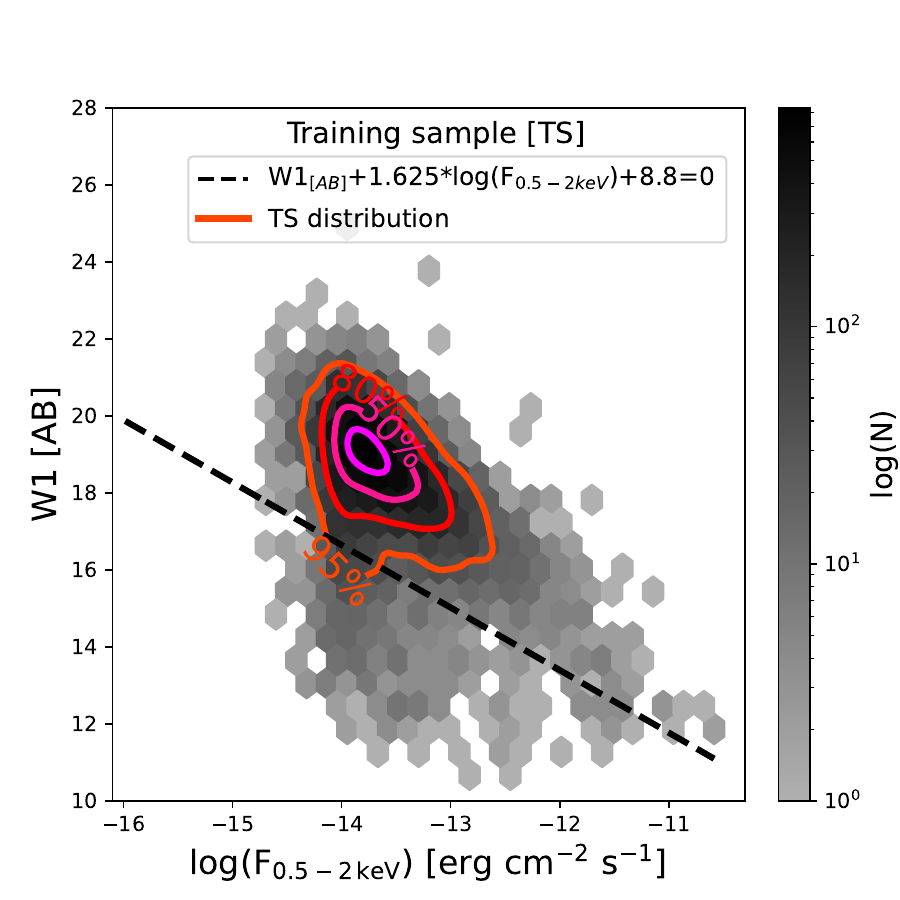}
\includegraphics[width=6cm]{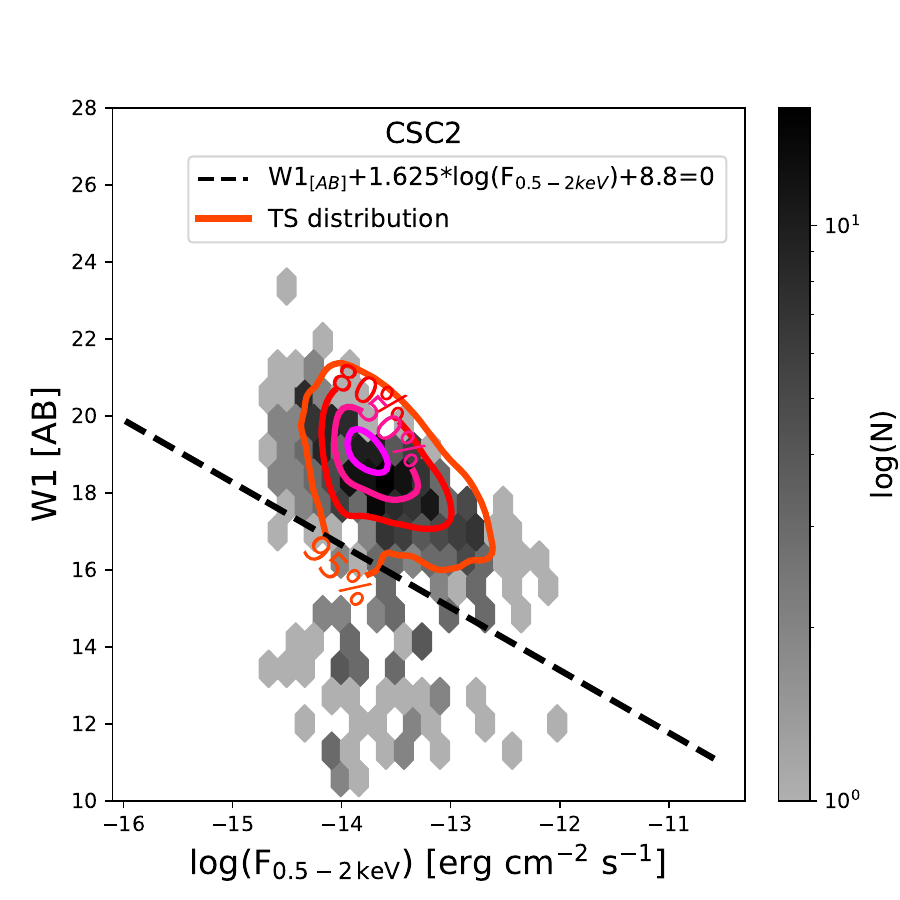}
\includegraphics[width=6cm]{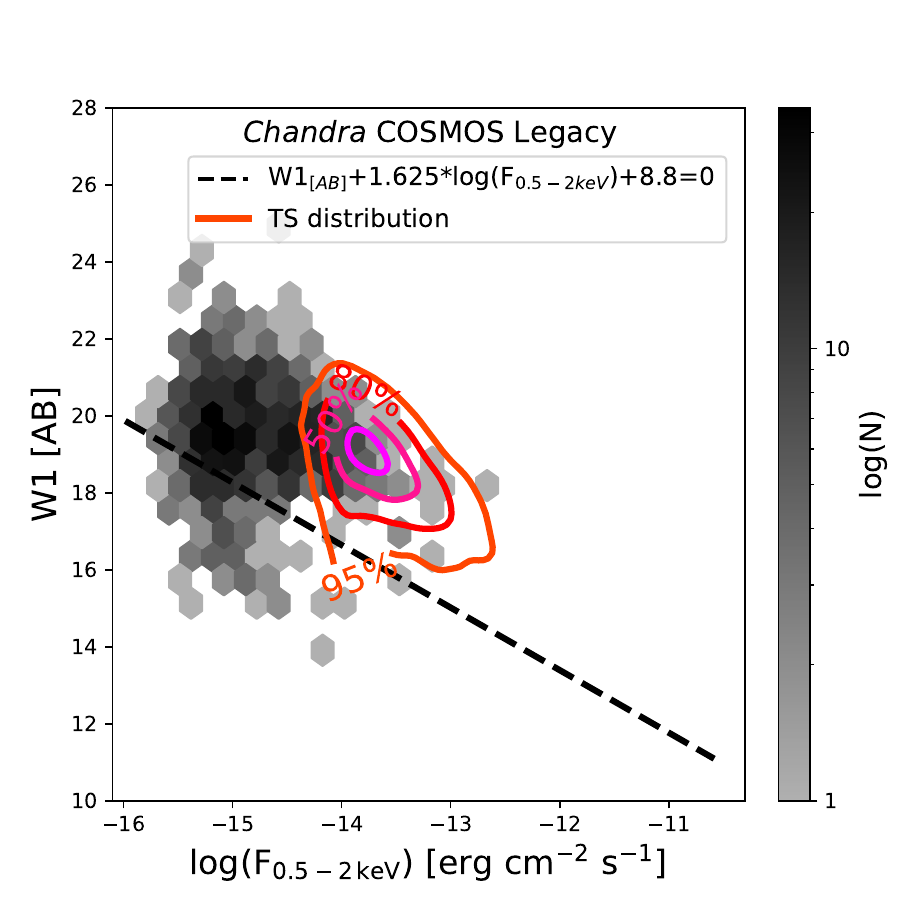}
\caption{X-rays flux vs. W1 magnitude distributions for the training (left panel), and the blind samples: CSC2 (middle panel) and {\it Chandra}-COSMOS (right panel). The dashed line represents the galactic/extragalactic separation proposed in \citep{Salvato18a}, who already pointed out that bright nearby galaxies would be misclassified. The contours encompassing 50\%, 80\%, and 95\% of the training sample distribution are over-plotted on all panels. While the CSC2 sample has a distribution similar to the training sample, the distribution of the {\it Chandra}-COSMOS Legacy sources is biased toward fainter values of X-rays and W1.}
\label{fig:XW1Comparison}
\end{figure*}

Table \ref{tab:breakdown} and Figure \ref{fig:redshift} show the final composition of the sample for training and blind samples as well as the source distribution in redshift, respectively. Figure \ref{fig:XW1Comparison} shows the distribution of the sources considered in this paper in the  X-rays flux and W1 magnitude plane, as defined in \citep{Salvato18a}. 
The redshift range of  CSC2 and {\it Chandra-COSMOS} are well represented by the sample used for training which matches CSC2 also in the W1 and X-rays flux of CSC2. Unlike CSC2, the bulk of the sources in the {\it Chandra}-COSMOS sample are much fainter than the training sample in both X-rays and W1 (see Figure \ref{fig:redshift}).

\section{The Method
\label{section:ARCH}}
This section describes our machine learning method and uncertainty quantification technique.

\subsection{Problem Formulation
\label{sec:problem_formulation}}

We denote the input data as $\mathcal{D}:= \{X,Y\}$, where $X \in \mathbb R^{M \times N}$ is the feature matrix with $M$ features for $N$ sources that we have discussed in Section \ref{subsec:features} and $Y \in \mathbb R^{N}$ represents the corresponding labels, which in our case, are given by spectroscopic redshifts. We denote the probability density function (PDF) of the normal distribution with mean $\mu$ and variance $\sigma$ as $p_{\mathcal N}(z; \mu, \sigma)$. A supervised learning approach to photo-zs would typically be tasked to find a parameterized function $f_\theta:\mathbb R^M \mapsto \mathbb R$, which, given photometry-based input features, predicts the redshift. The function may be a neural network, which typically is trained to minimize a loss function based on a training sample.

To illustrate our more general approach, we first consider a simpler case.
We assume that the conditional predictive distribution is represented by a Gaussian distribution with fixed, data-independent variance $\alpha^2$ and mean $y$ given by the neural network:
\begin{equation}
p(\hat y|x;\theta) = p_{\mathcal N}(\hat y; f_\theta(x), \alpha^2).
\end{equation} 
where $\alpha^2$ is the variance and  $\hat y$ is the predicted redshifts.

The free parameters of the network are then constrained by minimizing the Negative Log Likelihood (NLL) of predicting the spectroscopic redshift given the input photometry. This minimization problem is equivalent to minimizing the square of the  difference 
\begin{equation}
   L_2= \sum_{i=1}^{N-1} (\hat y_i - y_i)^2
\end{equation} 
between the predicted redshifts $\hat y$ and the ground truth $y$, a secure spectroscopic redshift. 
During inference, the mode (which in the case of Gaussian is also the mean) of the conditional predictive distribution is typically reported as the predicted solution. The variance of the Gaussian distribution remains constant regardless of the input data during the inference phase, leading to miss-calibrated uncertainty estimation. 
Moreover, a Gaussian probability distribution quantifying the redshift uncertainty is too restrictive. Due to potential color-redshift degeneracies \citep[e.g., Figure 1 in][]{Salvato18b}, the conditional predictive distribution is, in many cases, better described with a multi-modal distribution. 
To address this, \citet{D'Isanto18} adopted Convolutional Neural Networks (CNNs) combined with a mixture density network (MDN) to predict conditional predictive distributions based on multi-band imaging data.

We use a parameterized Gaussian mixture to describe the hypothesis class of the conditional predictive distribution, i.e.

\begin{equation}
    p(\hat y|x;\theta) = \sum_i^K \omega_i(x;\theta)p_{\mathcal N}(\hat y; \mu_i(x;\theta), \sigma_i^2(x;\theta)).
\end{equation}

However, as shown in \citet{D'Isanto18}, a single trained network producing a PDF cannot capture the uncertainty. We, therefore, extended our approach for uncertainty estimation and followed 
\cite{lakshminarayanan2017simple} who proposed to use an ensemble of deep neural networks such as MDNs to improve uncertainty estimates.
We follow this approach with an ensemble of MDNs that, given the photometric input features, each predicts the parameters of the Gaussian mixture distribution with a certain number of components. To be specific, the $i^{th}$ model $f_i(x;\theta_i)$ in an ensemble of $N$ models, given its parameters $\theta_i$ and the input features $x$, predicts the parameters of a Gaussian mixture composed of $n_i$ Gaussian, defined by the means $\mu^i(x; \theta_i)$, and standard deviations $\sigma^i(x; \theta_i)$ and weights $\omega^i(x; \theta_i)$. The conditional predictive distribution defined by this MDN is given by:
\begin{equation}
    p(\hat y|x, \theta_i)= \sum_{j=1}^{n_i} \omega^i_j(x; \theta_i) p_{\mathcal{N}}(\hat y; \mu^i_j(x; \theta_i)), \sigma^i_j(x; \theta_i).
\end{equation}
Finally, the predictions from all models are combined to give the posterior as  
\begin{equation}
\begin{aligned}    
    &p(\hat y|x, w_1, w_2, ..., w_N)= \\
    &\sum_{i=1}^{N} \sum_{j=1}^{n_i} \frac{\omega^i_j(x; w_i)}{N} p_{\mathcal{N}}(\hat y; \mu^i_j(x; w_i)), \sigma^i_j(x; w_i).
\end{aligned}
\label{eq:combination}
\end{equation}

For simplicity, the models are combined with equal weights. 
Each network in the ensemble is trained (see section~\ref{subsec:TTV}) independently, minimizing a suitable loss function (see section~\ref{subsec:loss}). 
The hyper-parameters of this approach include the number of trained models to combine and the number of Gaussian components for each model, $n_i$. The training conditions and model hyperparameters, including the number of Gaussians, are set differently for every network in the ensemble, and the training is described in section~\ref{subsec:hypersearch}. 
As shown in eq.~\ref{eq:combination}, the predictions of the individual models in the ensemble are combined equally to give the resulting conditional predictive distribution, which is also a mixture of Gaussians. The post-processing steps to summarize the resulting PDF, including identifying the mode, mean and uncertainties, are given in Section~\ref{subsec:postprocess}.
\begin{figure*}
\centering
\hspace*{-0.45cm} 
\includegraphics[width=19cm]{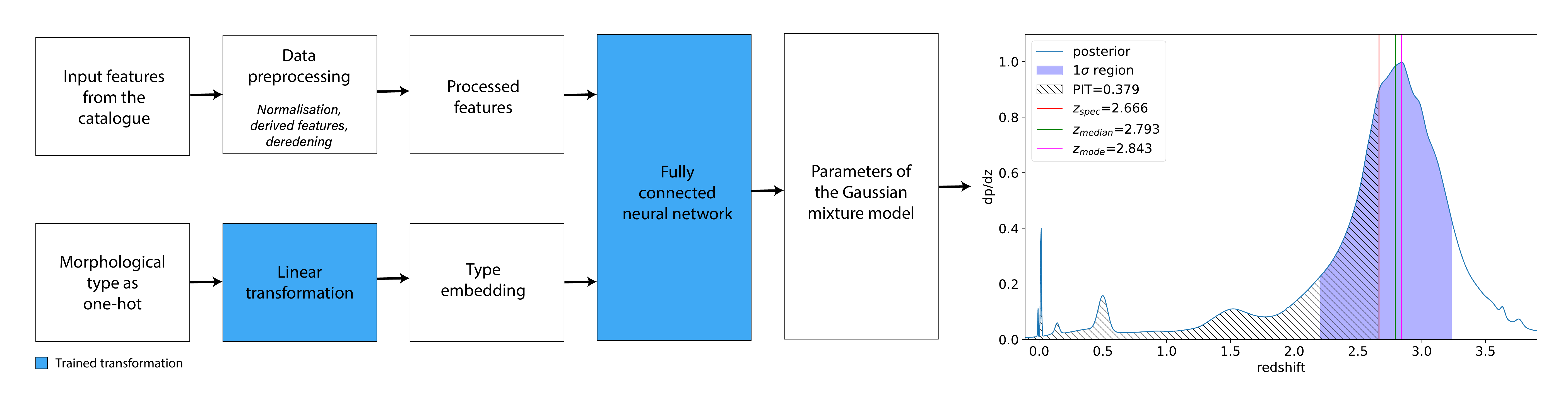} 
\caption{Training and inference pipeline. The blue boxes show the trained aspects of the model. We have intentionally chosen a complex posterior to demonstrate features. In most cases, posteriors produced by the model are close to a single Gaussian with a clear mono-modal prediction.}
\label{fig:architecture_pipeline}
\end{figure*}
\subsection{Training and inference pipeline
\label{subsec:TTV}}

We partition the samples into training, validation, and test splits roughly with proportions 80: 10: 10. We use the training split to train the model across different combinations of the hyperparameters. The model's performance on the validation split is evaluated to find suitable model architectures and training conditions, including parameter initialization, arguments for the optimizer, loss function, and others.
A visualization of the general pipeline that takes the input catalog and returns a parameter representation of the posterior is depicted in Figure \ref{fig:architecture_pipeline}. Among all features, we treat the vectorized morphological "TYPE" separately. Initially, this feature is transformed into a one-hot vector before applying a learnable linear transformation to embed them into $\mathbb R^n$, where we treat $n$ as a hyperparameter. For the rest of the features, we compute and store the mean and the standard deviation on the train and validation splits of the data and use them to normalize individual features. We utilize the identical mean and standard deviation values employed during the training phase for feature normalization during inference.

We employ a fully connected neural network to predict the distribution parameters from the concatenated feature representation.
In the following subsections, we discuss our choice of loss functions, training procedure, and post-processing the inference on new samples.
\subsubsection{Loss Functions
\label{subsec:loss}}

In the case of standard regression, we can find the network parameters as an approximation to the Maximum Likelihood Estimate (MLE) by minimizing the NLL loss. NLL loss can also be used for mixture density networks. Additionally, we implemented the Continuous Ranked Probability Score (CRPS) as done by \citep{D'Isanto18}. CRPS compares the Cumulative Distribution Function (CDF) of the predicted distribution against the Heavyside function, where the step is located at a value given by the ground truth, i.e., spectroscopic redshift. Let $F_x$ be the CDF of the conditional predictive distribution given the input $x$ and ground truth $y$; then the CRPS loss is defined as

\begin{equation}
    CRPS(F_x, y) := \int_{-\infty}^{\infty} (F_x(z) - \mathbbm{1}\{z - y\})^2 dz
\end{equation}

Let $p_\theta(y|x)$ be the predicted distribution and $q(y|x)$ represent the true conditional distribution\footnote{Assume that the data $\mathcal D := (X,Y)$ comes from some distribution $\hat q(x,y)$ with the corresponding true conditional distribution $q(y|x)$.}, given $y|x \sim q(y|x)$, the scoring rule $S(p_\theta, y|x)$ evaluates the quality of the predicted posterior in regards to the sample $y|x$. Higher values represent better alignment with the true posterior $q(y|x)$. The expected scoring rule concerning the true distribution can further be evaluated as 
\begin{equation}
    (p_\theta(y|x), q(y|x)) = \mathbb E_{y|x \sim q(y|x)}[S(p_\theta(y|x), y|x)].
\end{equation}
CRPS and NLL are proper scoring rules, i.e., optimal when the predicted conditional distribution is the same as the true conditional distribution. 

The latter can be shown to satisfy this property using Gibb's inequality as done in \citet{lakshminarayanan2017simple}, while the CRPS loss is shown to be a proper scoring rule in \cite{properScoringRuleandCRPS}. 

\subsubsection{Training procedure and hyperparameter search \label{subsec:hypersearch}}

The space of models that we considered is determined by the hyperparameters described in Table \ref{tab:modelhpm1}. We employed the normal Xavier initialization to initialize the network's parameters (i.e., weights), treating its parameters as hyperparameters. To find the most suitable network parameters, we minimize the loss functions described in the Subsection \ref{subsec:loss} together with the $L2$ regularization. We used the ADAM optimizer \citep[][]{kingma2017adam} along with a learning rate schedule. We randomly select hyperparameters concerning the model, optimizer, and scheduler for both loss functions for each configuration. We save the trained model if the corresponding loss on the validation samples is lower than a threshold. Upon randomly selecting the model architecture and training conditions, we gather $\sim$ $36$ models each while optimizing for NLL and CRPS losses. 
\begin{table}[h]
    \centering
    \caption{Model and training hyperparameters with the range within which they are selected.}
    \begin{tabular}{l c}
\hline
        Hyperparameter & Allowed range \\
        \hline
        \hline
        Hidden Layers & $[11, 14]$ \\
        Neurons per layer & $[110, 135]$ \\
        Dropout & $0.0$ \\
        Number of Gaussian & $[2,4,6,8]$ \\
        Type embedding dimension & $10$ \\
        Gain(model initialization) & $[0.85, 1.05]$ \\
        Learning rate & $[0.00005, 0.005]$\\
        Weight Decay & $[0.0000005, 0.003]$\\
        Step size (StepLR Pytorch's scheduler) & $[20]$\\
        Gamma (StepLR Pytorch's scheduler) & $[0.1, 0.5]$\\
        \hline
    \end{tabular}
    \tablefoot{These specific ranges are selected by observing model performance with hyperparameter values in wider ranges.}
    \label{tab:modelhpm1}
\end{table}

\subsubsection{Inference and post-processing
\label{subsec:postprocess}}

To predict the distribution for new samples, we combine the models' results uniformly as discussed in Section \ref{sec:problem_formulation}. We compare the results evaluated on the validation and test merged samples for the combination of the models trained on CRPS loss and those trained on NLL loss in Table \ref{tab:modelhpm}. We performed slightly better by combining CRPS and NLL models than by considering models trained on individual loss functions (see Table \ref{tab:modelhpm}). The resulting posterior of the ensemble has 184 Gaussian (GMM) in the mixture.

To obtain a point estimate of the redshift, we find the mode of the resulting Gaussian distribution because the mean or median could be misleading for bi- or multi-modal distributions. In addition to the point estimates, we provide the redshift posterior for all sources computed as
the sum of the Gaussian distributions in the GMM. We provide this in the form of a set of posterior values for 2000 points linearly spaced in the range ($0<z<=7$) \footnote{We also provide the GMM values so the user can recompute on a different grid if preferred}. We also provide the upper and lower 1$\sigma$ and 3$\sigma$ errors from the posterior. We do this by finding the values where the cumulative distribution function is equal to the equivalent 1$\sigma$ and 3$\sigma$ probabilities of a normal distribution (see the left panel of Figure \ref{fig:architecture_pipeline} for a schematic representation). 
\begin{table*}
    \centering
    \caption{Averaged statistics across all individually trained NLL loss and CPRS models.}
    \begin{tabular}{cccccc}
    \hline
        Ensembles & Num. Gaussians & CRPS loss & NLL loss & Fraction of outliers & Accuracy  \\
        & & & & $\eta$&$\sigma_{\rm NMAD}$ \\
        \hline
        \hline
        Models trained on CRPS loss & $132$ & $0.145$ & $-0.295$ & $0.121$ & $0.068$ \\
        Model trained on NLL loss & $156$ & $0.146$ & $-0.383$ & $0.12$ & $0.065$ \\
        All combined & $288$ & $0.143$ & $-0.361$ & $0.116$ & $0.066$ \\
\hline
    \end{tabular}
    \tablefoot{the table includes aggregated information for the combined ensemble of all NLL and CRPS models. Furthermore, the fraction of outliers and accuracy are listed for the best-performing ensembles.}
    \label{tab:modelhpm}
\end{table*}
\section{Redshift computation for eFEDS sources and comparison with literature
\label{sec:eFEDSquality}}

\begin{figure}
\centering
\includegraphics[width=0.9\columnwidth]{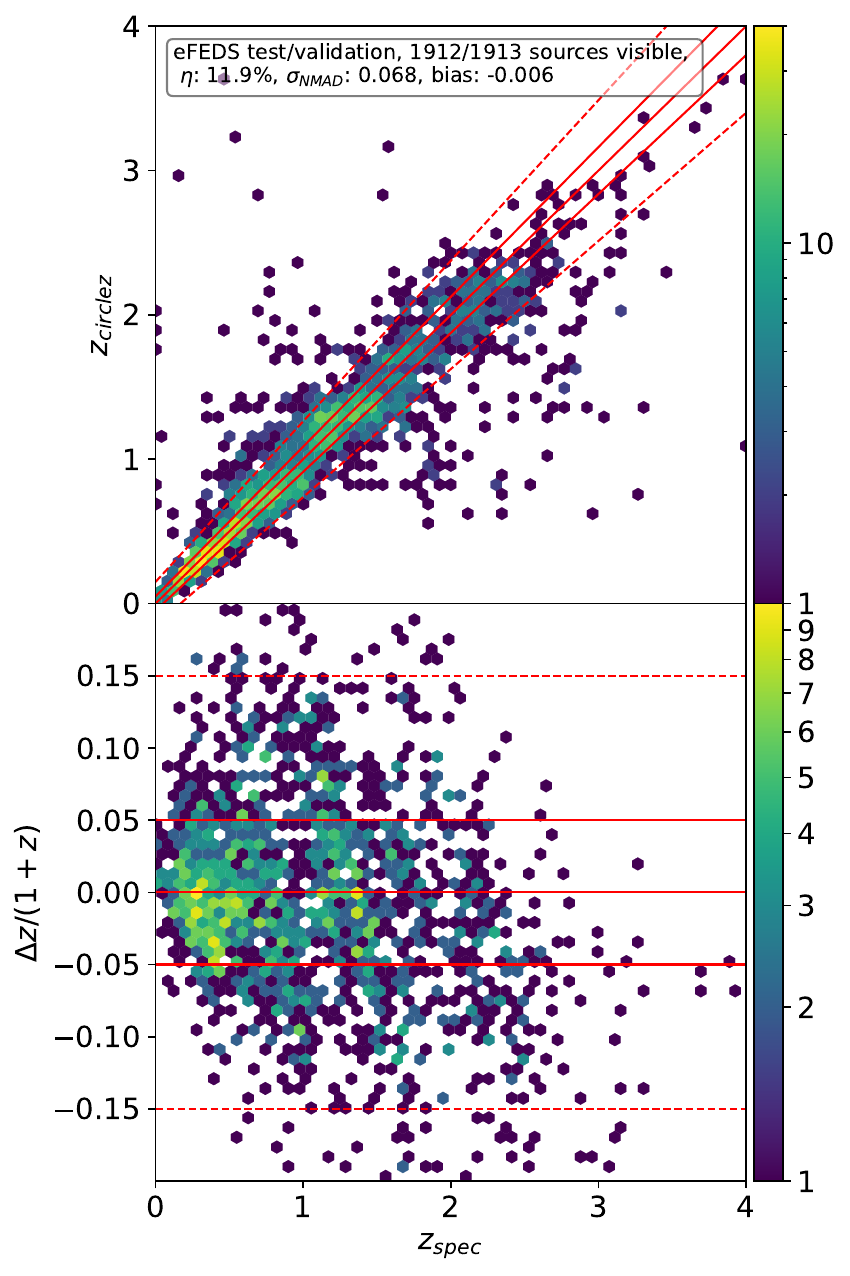}
\caption{Comparison between photo-z computed with {\sc{CircleZ}}  and spectroscopic redshifts in eFEDS, considering only the test and validation sources.}
\label{fig:ZZplot_z_spec_z_circlez_eFEDS}
\end{figure}

Using the model prediction described in Section \ref{section:ARCH} we have extended the computation of redshift to the 22079 sources in the eROSITA/eFEDS catalog of counterparts \citep{Salvato22} that have been classified as extragalactic (CTP\_CLASS>2), using only LS10-South data. For the eFEDS sources in the test and validation sample the comparison between the spectroscopic redshift and the photo-z computed with  {\sc{CircleZ} is shown in Figure \ref{fig:ZZplot_z_spec_z_circlez_eFEDS}}. 
 As a reminder, \citet{Salvato22} presented two different methods of computing photo-zs, one based on SED fitting {\sc{LePHARE}} and one based on machine learning {\sc{DNNZ}}, the latter also based on CNN and with an architecture very similar to the one of {\sc{CircleZ}}. In particular, in \citet{Salvato22}, the metrics (accuracy and fraction of outliers) were split into three distinct regions based on the photometry quality.
Region 1, centered on the GAMA09 field, has the best data, including optical data from HSC, optical and near-infrared data from KiDS-Viking, optical and Mid-infrared data from LS8, and UV from GALEX. This is the region for which  {\sc{LePHARE}}  performed best. In Region 2, there is a drop in accuracy and an increase in the fraction of outliers due to the lack of deep near-infrared data and probably some issue with the calibration of the {\it r} and {\it i} band in HSC. The latter is suggested by the fact that {\sc{DNNZ}} follows a similar trend and has a drop in the accuracy and the fraction of outliers in Region 2, as {\sc{LePHARE}}, despite relying only on HSC data. Region 3 is the region where data are very pure in general and not considered further in the comparison.

The metrics (this time also including bias) for Regions 1 and 2 for {\sc{LePHARE}} and {\sc{DNNZ}} are reported in Table \ref{tab:accuracyBreackdown} for comparison, together with the same metrics computed using {\sc{CircleZ}} for the sources in eFEDS that are part of the training, test, validation samples, and overall.
By comparing the metrics, it is clear that the new photo-z are superior to those previously available for that area, despite relying only on optical and mid-infrared data from LS10. The accuracy is slightly worse than for {\sc{DNNZ}}, in general, and {\sc{LePHARE}} in Region 2. However, the fraction of outliers has noticeably dropped for all samples. The bias is also slightly worse than for {\sc{DNNZ}} and {\sc{LePHARE}} in Region 1, but comparable for Region 2. Region 1 is the area with the reacher, deeper, and more homogenous photometry  from KiDS and HSC \citep[see][for details]{Salvato22}.

It is also worth mentioning that the metrics reported for {\sc{LePHARE}} and {\sc{DNNZ}} were measured using the training sample, not on the test and/or validation sample.
If we compute the metrics for {\sc{CircleZ}} using all the spectroscopy, i.e., including those used for the training and validation, the values would suggest a quality close to the one available for the best pencil-beam surveys \citep[XMM-COSMOS, (E)CDF][]{Salvato09, Luo10, Hsu14}, computed with deep, rich and homogenized photometry).  This assures that the new photo-zs made available for eFEDS in this paper are reliable and can be used for Luminosity Function studies or for preselecting specific instruments or wavelengths of interest when submitting a proposal.

\begin{figure}
\centering
\includegraphics[width=10cm]{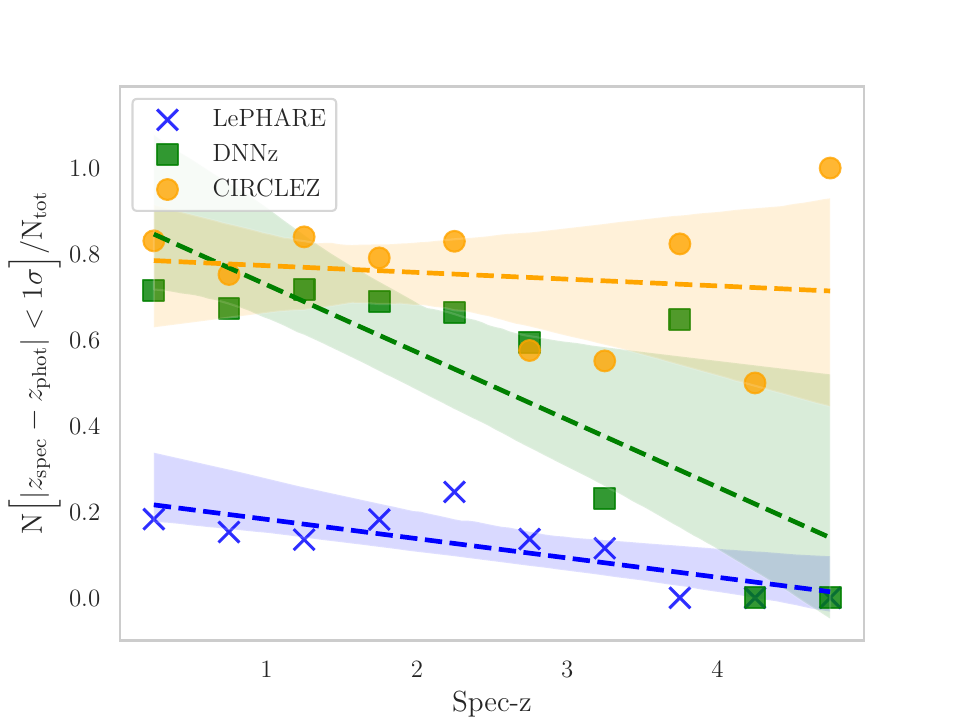}
\caption{Fraction of sources with spectroscopic redshift within 1$\sigma$ from the photo-zs, in eFEDS. In addition to the trend per bin (crosses, filled dots, and squares), we plot the regression fits (dashed line) as well as 95\% confidence intervals (shaded regions).}
\label{fig:frac_1_sigma}
\end{figure}

The new photo-zs also provide more realistic PDZ,  1 and 3$\sigma$ errors. In fact, it is well known that SED fitting tends to underestimate the errors. Usually, the fraction of sources with spectroscopic redshift within 1$\sigma$ from the photometric ones is very low \citep[e.g.,][]{Dahlen13}. The area within the photo-zs and $/pm$ 0.1 of the PDZ (PDZ\_BEST in {\sc{LePHARE}}) is usually taken as a measure of the reliability of the photo-zs. When PDZ\_BEST is high, the solution is very peaked, while a low PDZ\_BEST indicates large uncertainties and degeneracy of the solutions. In \citet{Brescia2019}, it was shown that for the AGN in STRIPE82-X \citep{Ananna2017}, the PDZ\_BEST was high for many outliers.  Figure \ref{fig:frac_1_sigma} shows this fraction in bins of redshift for {\sc{LePHARE}}, {\sc{DNNZ}}, and {\sc{CircleZ}} applied to eFEDS. {\sc{CircleZ}} and {\sc{DNNZ}} perform better than {\sc{LePHARE}}, with {\sc{CircleZ}} performing better overall at z>2.  

\begin{table*}
\centering 
\caption{Comparison of bias, fraction of outliers, and accuracy between photo-z computed with {\sc{CircleZ}}, {\sc{LePHARE}}, and {\sc{DNNZ}}, for extragalactic sources in the eROSITA/eFEDS field.} 
\small
\begin{tabular}{lcccc}
\hline
\\
 Survey & Number of sources &bias & accuracy & fraction of outliers \\
  & & & $\sigma_{NMAD}$ & $\eta$ [\%]\\
 \hline
 \hline
 \\
{\sc{LePHARE}} Region 1   & 3472       &  0.003  & 0.049 & 14.1  \\ 
{\sc{DNNZ}} Region 1  &  3419        & -0.001  & 0.039 & 16.7  \\ 
{\sc{LePHARE}} Region 2   &1884        &  0.006  & 0.068 & 23.8  \\ 
{\sc{DNNZ}} Region 2    & 1808         & -0.004  & 0.052 & 27.2  \\ 
\\
\hline
\\
{\sc{CircleZ}} on test\&validation & 2913  & -0.005 & 0.067 & 11.6   \\ 
{\sc{CircleZ}} on  training   & 11651 &    -0.004 & 0.049 & 7.2   \\
{\sc{CircleZ}} on training, test, and validation combined  &14564     &-0.004 & 0.055 & 8.6    \\ 
\\
\hline
\\

{\sc{CircleZ}} on eFEDS test\&validation & 1913 & -0.006 & 0.068 & 11.9   \\ 
{\sc{CircleZ}} on eFEDS training   & 8032 &     -0.004 & 0.052 & 7.8   \\ 
{\sc{CircleZ}} on eFEDS all       &   10044     &-0.005 & 0.052 & 8.1    \\ 
\\
\hline
\\
{\sc{CircleZ}} on CSC2 & 416 &-0.009 & 0.055 & 12.3 \\
{\sc{CircleZ}} on {\it Chandra}-COSMOS Legacy &1699& -0.041 & 0.124 & 32.4 \\  
{\sc{CircleZ}} on {\it Chandra}-COSMOS Legacy {\it i} mag<21.6 & 813 & -0.010 & 0.067 & 10.1     \\ 
\\
\hline
\end{tabular}
\tablefoot{The values for {\sc{LePHARE}} and {\sc{DNNZ}} are taken directly from \citep{Salvato22} for comparison. While the selection of the templates for {\sc{LePHARE}} is based on the same sample used for assessing the quality of photo-zs, the training for {\sc{DNNz}}({\sc{CircleZ}}) is partially(completely) independent from the test and validation sample used here. The quality of {\sc{CircleZ}}  exceeds the one obtained by both {\sc{LePHARE}} and {\sc{DNNZ}} in the area outside KiDS, demonstrating the strength of the method in general and particularly where there is a lack of near-infrared photometry.}
 \label{tab:accuracyBreackdown}      
 \end{table*}
%

section{New photo-z catalog release for eFEDS \label{sec:DataRelease}}
With this paper, we release a new catalog of photo-z computed with {\sc{CircleZ}}, for the extragalactic sources in the eROSITA/eFEDS field. For each source we provide, in addition to the eROSITA\_ID, the photo-z, and the 1 and 3 $\sigma$ lower and upper errors,  also a unique identifier for the counterpart in the LS10 catalog. The identifier is created by concatenating the columns RELEASE, BRICKID, and OBJID. In addition, we provide about 5000 new spectroscopic redshifts that became available with the 18th Data Release of SDSS-V \citep{Almeida2023}. The catalog is available via Vizier and on the eROSITA/EDR page\footnote{\url{https://erosita.mpe.mpg.de/edr/eROSITAObservations/Catalogues/}. The redshift probability function (PDZs) for each source is also available, under request.}

\section{Robustness tests
\label{sec:comparison}}

\begin{figure*}
\centering
\includegraphics[width=0.67\columnwidth]{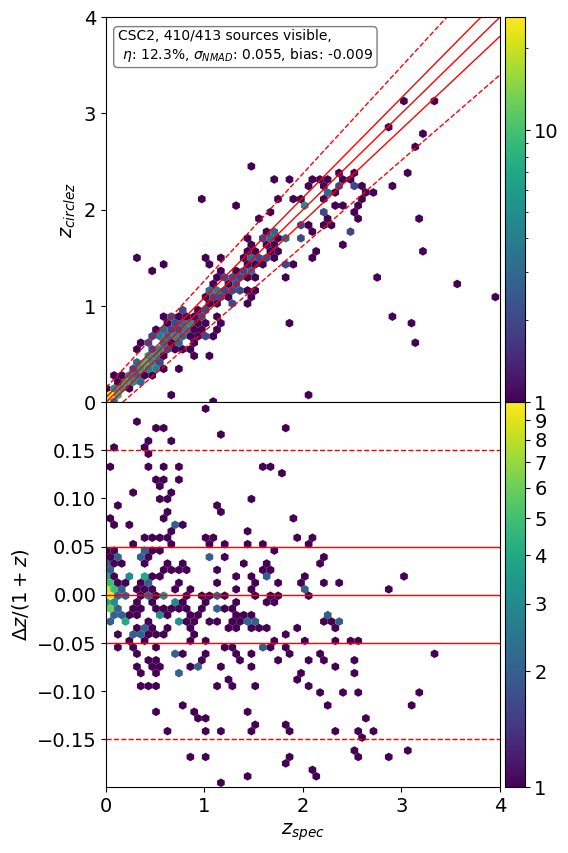}
\includegraphics[width=0.67\columnwidth]{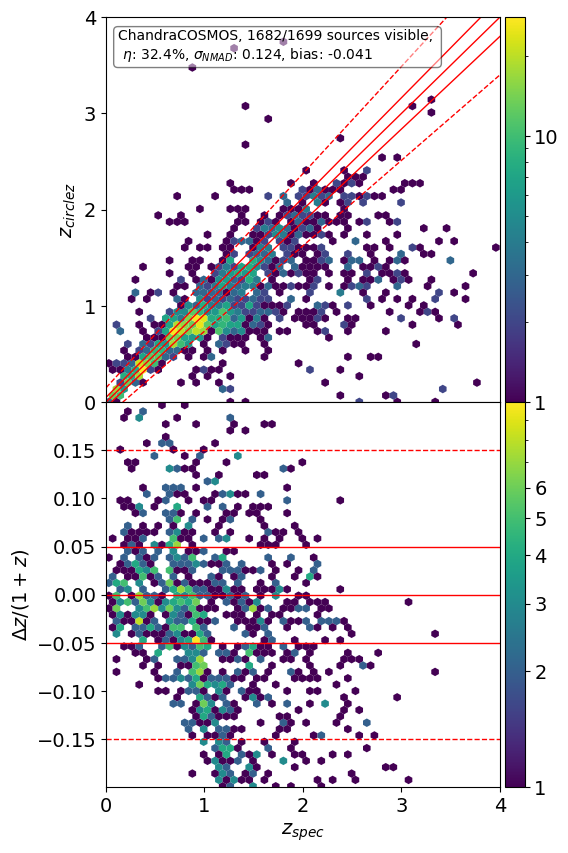}
\includegraphics[width=0.67\columnwidth]{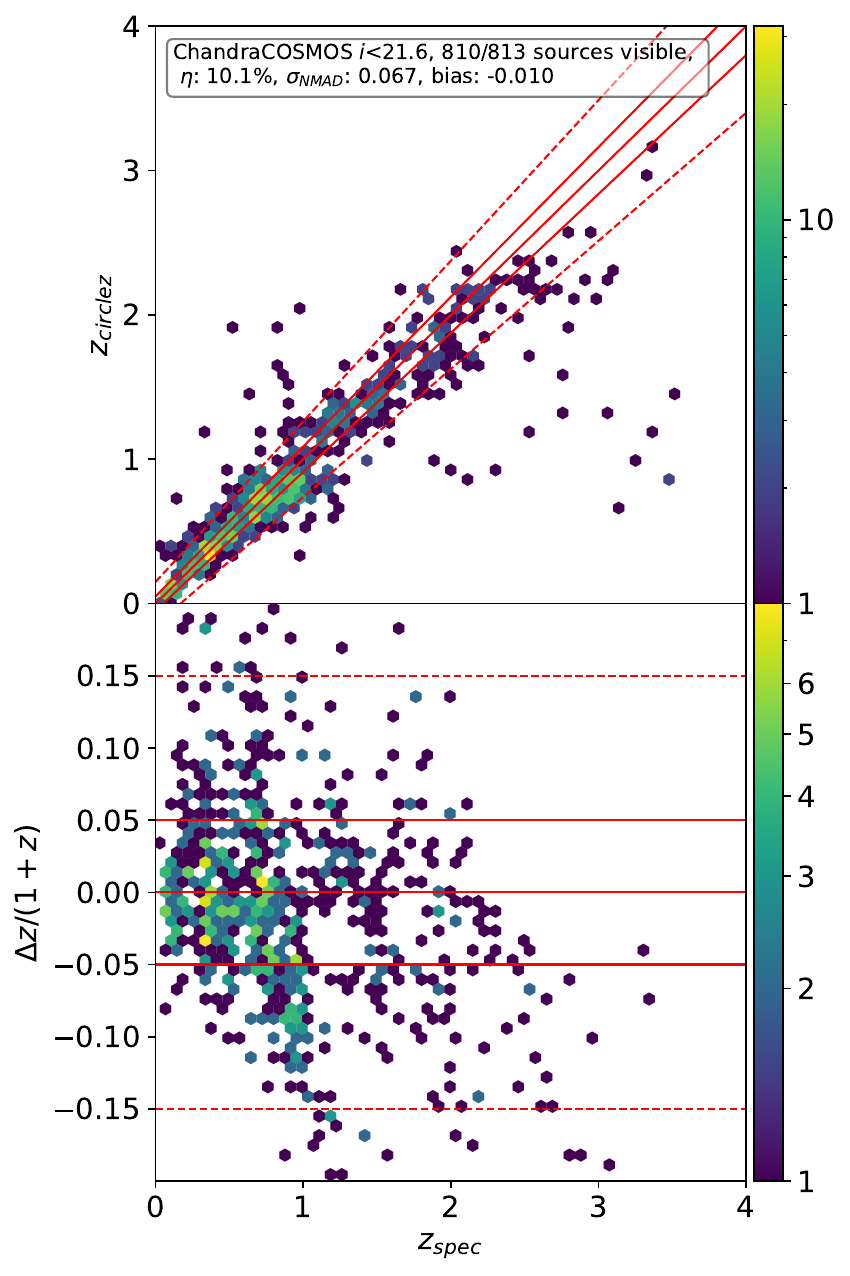}
\caption{{\sc{CircleZ}} photo-zs vs spectroscopic redshifts for the CSC2 sample (left panel) and {\it Chandra}-COSMOS Legacy  sample from \citep[][]{Marchesi16} (middle and right panel). Numbers in the figures show the number of objects visible in this redshift range instead of the full sample number. In {\it Chandra}-COSMOS, the performance is reduced due to the increased faintness of the survey and, therefore, fainter counterparts.}
\label{fig:ZZplots}
\end{figure*}

Almost 60\% of the training samples' sources are from eFEDS, where the X-ray observations and the LS10 photometry are homogeneous. Then, one could argue that the quality of the photo-z is field-dependent. To check the robustness of {\sc{CircleZ}} we computed photo-z for the two blind samples 
presented in  Section \ref{subsec:BlindSamples}). The results on various metrics are reported in Table \ref{tab:accuracyBreackdown}.
First, we applied {\sc{CircleZ}} on the CSC2 sample, which has a distribution in X-ray and W1 flux similar to the training sample. Unsurprisingly, $\eta$ and $\sigma_{\rm NMAD}$ are comparable to those obtained for the training and validation sample (see left panel of Figure \ref{fig:ZZplots}).   

The middle and right panels of the same Figure show the performance on the {\it Chandra}-COSMOS sample of \cite{Marchesi16}. As expected, the performance for this faint sample is dramatically reduced compared to the X-rays brighter eFEDS sample, given that the counterparts for these fainter X-ray sources will also have faint photometry and thus lower signal-to-noise in the shallow LS10 survey. 
However, if we limit the sample to sources brighter than about 21 magnitudes (see Table \ref{tab:accuracyBreackdown}), the quality of the photo-z is consistent with previous results, suggesting that we are limited only by the depth of the data.\\

\subsection{PIT score
\label{subsec:PIT}}

\begin{figure}
\centering
\includegraphics[width=0.65\columnwidth]{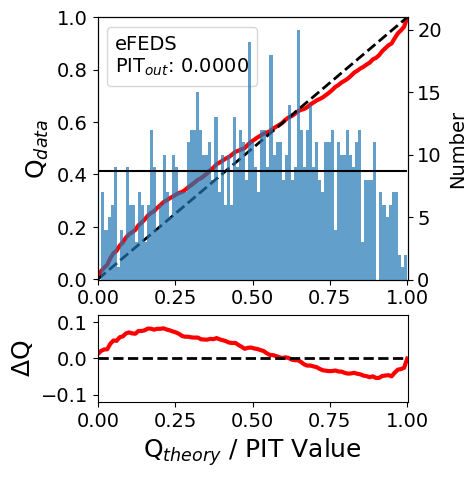}
\includegraphics[width=0.65\columnwidth]{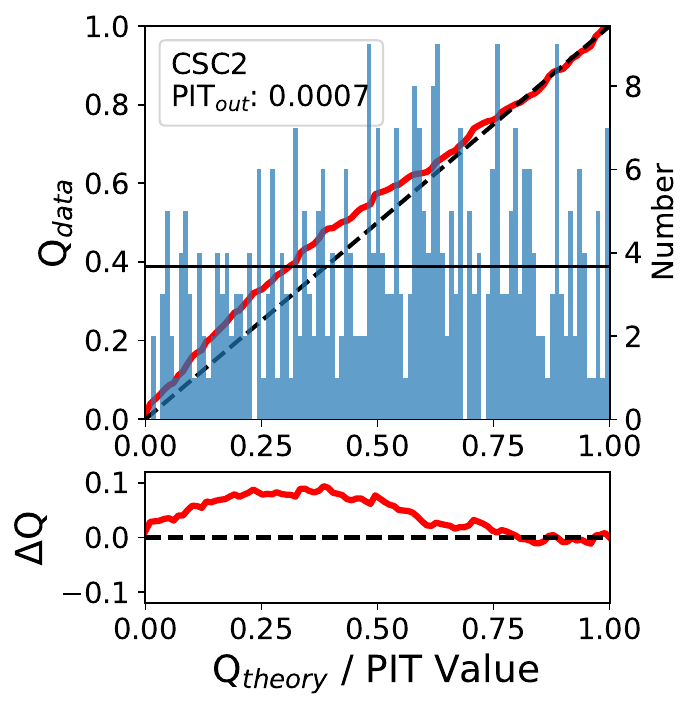}
\includegraphics[width=0.65\columnwidth]{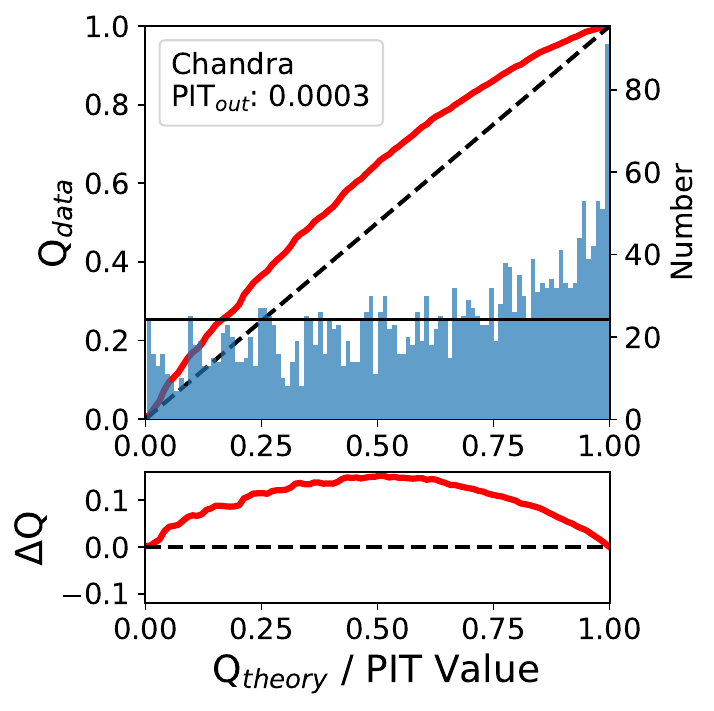}
\includegraphics[width=0.65\columnwidth]{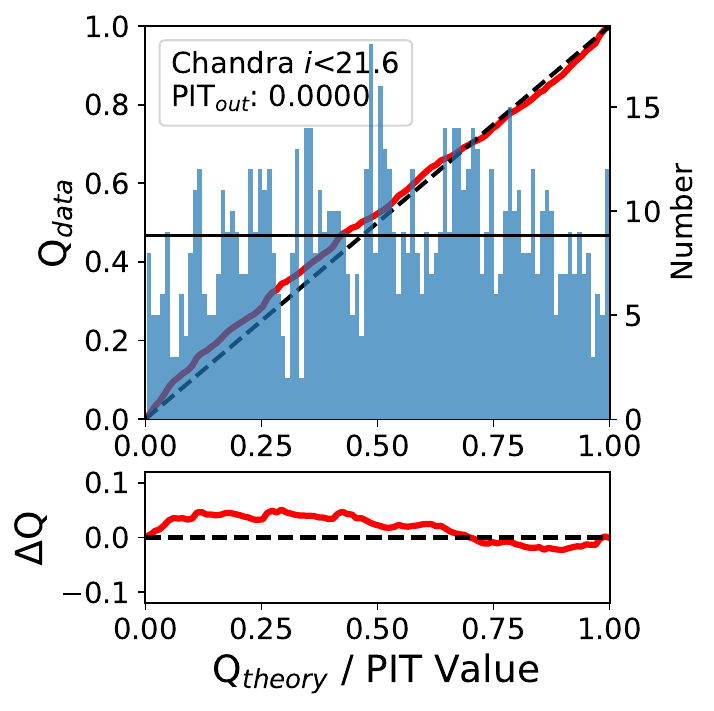}
\caption{PIT distribution and quantile-quantile plot between photo-z computed with {\sc{CircleZ}} and the spectroscopic redshifts for the test and validation sources in eFEDS (top panel), CSC2 (second row) and {\it Chandra}-COSMOS Legacy without (third row) and with a cut in magnitude (bottom panel). While the first two samples perform well, the PIT distribution for {\it Chandra}-COSMOS is in excess of the upper end.}
\label{fig:PITs}
\end{figure}

For scientific exploitations, in addition to a point prediction, reliable errors must be provided by any photo-z tool. In particular, the full posteriors to reflect our knowledge of the redshift is needed, for example, in the computation of luminosity functions \citep[e.g.,][]{Buchner15, Fotopoulou16, Miyaji15}. For this reason, we also looked at the Probability Integral Transform (PIT:\citet{D'Isanto18,Schmidt2020}) for all three samples. 
The PIT value is the integral of the posterior up to the spectroscopic redshift (see left panel of Figure \ref{fig:architecture_pipeline}) and would be uniformly distributed if the spectroscopic redshifts were drawn from the posteriors. The results for the test sample in eFEDS, CSC2, and {\it Chandra}-COSMOS are shown in Figure \ref{fig:PITs}. The quantile-quantile plots show the PIT value against a uniformly distributed value.  The figures were produced using the Redshift Assessment and Infrastructure Layers (RAIL) framework, discussed in detail in \citet{Schmidt2020}. 
The Figure demonstrates the overall good performance of the posterior distribution and the point prediction performance. It demonstrates that the full posteriors are neither overly nor underly dispersed for eFEDS and CSC2. If the spectroscopic samples were drawn from the posterior, the PIT distribution would be uniform. An excess of PIT values at the lower or upper ends shows that the posteriors are not identifying true values at their lower or upper ends, as it is happening for {\it Chandra}-COSMOS. An excess in the center shows that the posteriors are underdispersed and underestimate the errors on the point predictions.

\subsection{Limitations on photo-zs for all-sky surveys \label{subsec:limitations}}

\begin{figure}
\centering
\includegraphics[scale=0.45]{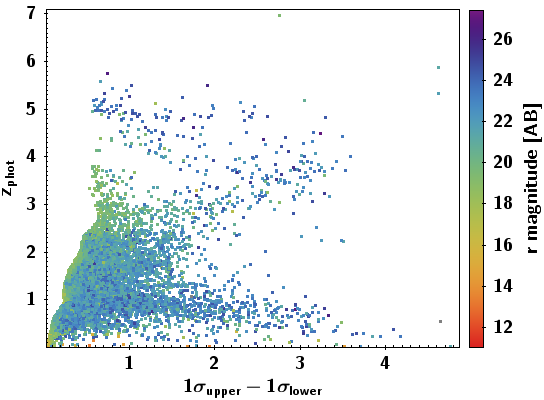}
\caption{Photo-z vs 1$\sigma$ error for all the sources in eFEDS with spectroscopic redshift available, color-coded by their magnitude in the r band.}
    \label{fig:photoz_1sigma}
\end{figure}

To further test the robustness of {\sc{CircleZ}}, in Figure \ref{fig:photoz_1sigma}, we plotted the photo-z computed in eFEDS and the width of the 1 $\sigma$ error,  color coding the sources as a function of their magnitude. The scatter plot is broad, but a sharp line delimits the smaller error associated with the photo-z, corresponding to the brighter sources. The brightest sources typically also have smaller photometry uncertainties (and thus high SNR) and, thus, more accurate photo-zs. As it is visible in Figure \ref{fig:LS10Depth}, the depth of LS10 is not homogeneous, implying that the quality of the photometric redshifts for all-sky surveys, such as eROSITA, will have various grades of accuracy. However, the PIT score analysis on CSC2 (random distribution in the sky) demonstrated that the PDZs correctly account for this limitation.

\section{Discussion 
\label{sec:discussion}}

Thus far, there has been scant optimism regarding the feasibility of attaining accurate photo-zs for AGN in all-sky surveys.
This was due to the assumption that many photometric bands from different surveys needed to be assembled and cross-calibrated  (unthinkable to be achieved for wide areas). While it remains true that more bands better describe the SED of these complex sources, in this work, we have demonstrated that using an adequate training sample and going beyond the use of fluxes and colors with a machine learning approach,  reliable photo-zs can be achieved for the first time, using only optical and WISE data as offered by LS10-South.
In particular,  the features include information on the light distribution of the sources, parameterized by the fluxes and colors for different apertures, and the quality of the fit assuming different morphological types, in addition to photometric errors and tractorized \citep{Lang14} WISE photometry. 

Currently, LS10 is the only survey that provides this kind of information,  and we argue here that surveys such as the Legacy Survey of Space and Time \citep[LSST;][]{Ivezic2019} and {\it Euclid} \citep[][]{Mellier2024} should provide the same kind of output.

The {\it Vera C. Rubin} Observatory will utilize the pipeline already tested on Hyper Suprime Cam and presented in \cite{Bosch2018}. The plan\footnote{\url{https://lse-163.lsst.io/}} is for the pipeline to provide fluxes for several concentric apertures, together with total flux and errors assuming two best-fitting models: a PSF and a composite bulge+disk model. The multi-epoch observations with LSST will also help improve photo-zs for AGN in two ways. First, a correction for variability could be applied to the photometry as already successfully implemented in  \citet{Salvato09}. Second, variability is at least in part also redshift-dependent (Satheesh-Sheeba et al., in prep).

{\it Euclid} would offer further constraining power due to its exceptional spatial resolution and wavelength coverage extending to near-infrared photometry. However, no plans exist to provide aperture photometry in multiple annuli as part of the standard data release products \cite{Merlin2019}. \cite{Guy22} recommends the joint processing of {\it Rubin} and {\it Euclid} data products, including aperture fluxes and other measures. A high density of apertures could significantly impact photometric redshift performance using the method presented here and add significant value to the resultant data sets. 
However, there is currently no plan for both surveys to combine their photometry with WISE. At the same time, the analysis of the features presented in Table \ref{tab:Marianna} suggests that the features from WISE are key in all analyzed subgroups.

\subsection{Outliers as a function of neighbors \label{subsec:neighbors}}
\begin{figure}
\centering
\includegraphics[width=9.1cm]{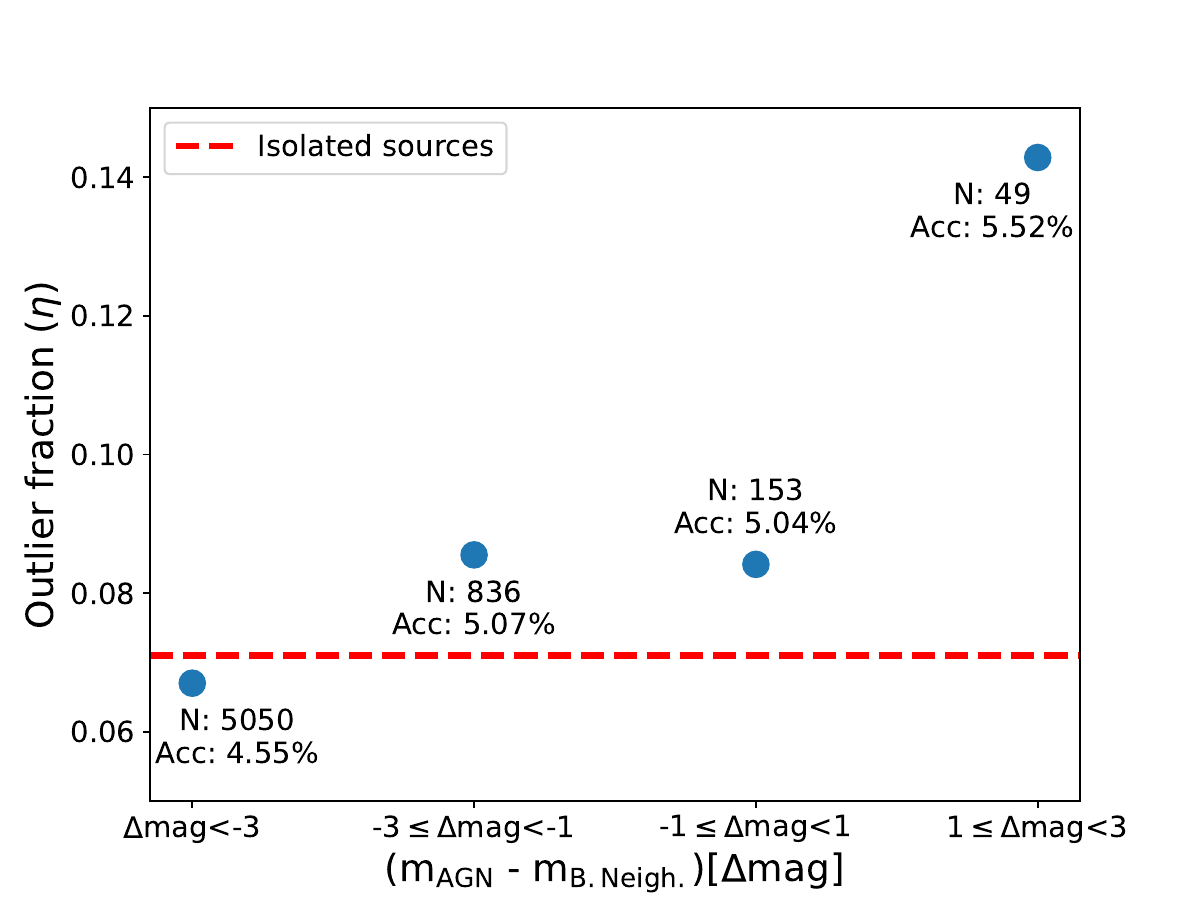}
\caption{Outlier fraction as a function of binned relative neighbor brightness. The outliers increase in the presence of bright neighbors.} 
\label{fig:neighbors}
\end{figure}

It is important to stress that there is room for improvement in what LS10 provides. 
As measured by LS10, the photometry within apertures does not account for the additional flux emitted by neighbor sources. We tested whether this could explain at least some of the outliers. For that, we identified the neighbors within 5" from all eFEDS sources with spectroscopy. We then considered the brightest of the neighbors and the magnitude difference from the central source ($\Delta mag$).
We categorized the neighbors into classes based on their relative brightness: "isolated" (no neighbor), "more than 3 magnitudes dimmer," "between 3 and 1 magnitude dimmer", "between 1 magnitude dimmer and 1 magnitude brighter," "between 1 and 3 magnitudes brighter" and "more than 3 magnitudes brighter". These criteria must be met in all 4 optical bands {\it g, r, i} and {\it z}. Imposing the same trend in all bands reduces the sample that can be analyzed, but in return, we get a clearer correlation.
In Figure \ref{fig:neighbors}, we report the mean value of the outlier fraction for the sources in various bins of $\Delta mag$, indicating with a red dashed line the outliers fraction for the isolated eFEDS sources sample. Sources with no neighbors or very faint neighbors have a lower fraction of outliers, which increase with the increasing brightness of the neighbors.\\
This implies that the quality of the photo-zs could improve further if the Imaging Legacy Survey for DESI would consider providing a segmentation map for redefining the area to be used for measuring the aperture fluxes \citep[e.g., SeXseg,][]{Coe2015}in their future data releases. More about this will be discussed in Roster et al. (in prep), where photo-zs for AGN are computed using directly the images.

\section{Conclusions and Summary
\label{sec:Conclusions}}

In this paper, we present {\sc{CircleZ}}, a new algorithm for estimating redshifts for AGN. The model is based on FCNN\footnote{Fully Connected Neural Network}, applied to aperture photometry from optical to mid-infrared data from LS10-South, already publicly available over the entire footprint of eROSITA-DE.
 For this purpose, we have constructed a sample of 14173 sources of X-ray-detected AGN enriched by a sample of 391  AGN at high-z to be used as training, test and validation samples (see Section \ref{SEC:training}). We did not limit the features to those expressing total fluxes (together with their errors) and colors, as is done usually in SED fitting. Instead, we considered all features that are, at least at some level, redshift-dependent (see Section \ref{sec:FeaturesRelevance}).  These features are related to the light profile of the galaxies and the apparent host/AGN relative contribution and are described in Section \ref{sec:Data} and Figure \ref{fig:LS10Features}. In short they are fluxes within various apertures, corrected for Galactic extinction, from which also colors within annuli are computed. In addition, we considered the quantity "dchisq" which expresses how closely the azimuth profile of a source resembled the PSF, DEV, REX, and SER model profiles. All the features are either already offered in the standard  LS10 catalog or easily computed from the available columns.
 
The model obtained in the training was then applied to the eROSITA/eFEDS field (Section \ref{sec:eFEDSquality}). Overall, we have obtained a comparable or better quality photo-z to what was previously available in the best region of the field (REGION 1), with much less effort than what was described in \citet{Salvato22}, despite lacking UV, u and  and Near-Infrared band photometry.
With the paper, we release the new catalog of photo-zs in eFEDS, including 1 and 3 $\sigma$ errors and PDZ (Section \ref{sec:DataRelease}).

We tested the stability of the results (Section \ref{subsec:BlindSamples}) by using {\sc{CircleZ}} and the model obtained in the training phase on two additional samples, used for a blind test. The first sample, formed by 651 sources from CSC2 in the footprint of LS10-South, shares the same properties (in terms of WISE and X-ray fluxes) as the training sample. There, the quality of the photo-zs is consistent with the one obtained in the validation sample limited to eFEDS sources. The second blind sample consists of all the AGN detected in the {\it Chandra}-COSMOS Legacy survey, with reliable spectroscopic redshift and detected in t LS10-South. The metrics indicate that the quality of the photo-zs is consistent with the one computed in the test sample of eFEDS and in CSC2, as long as we consider a limiting AB magnitude of 21.7 in {\it i} band. The accuracy degrades rapidly for fainter sources (and thus with larger photometric errors).  

This implies that the method is stable and reliable, and what is now needed for computing reliable photo-zs for AGN in all-sky surveys is photometry computed as in LS10 but at a deeper depth over the entire sky. Future deep photometric surveys such as LSST and {\it Euclid}, given that they will allow us to identify the faint and high-z AGN, should consider providing the same quantities provided by LS10, including the matched and de-blended photometry to the WISE data.
More information on morphological and spatial properties will improve not only photo-zs but also the classification and identification of AGN as was already demonstrated in \citet{Doorenbos2022, Savic2023}.
Ideally, Both surveys should provide the photometry in various apertures, after masking nearby objects (this is a current limitation in the data provided by LS10).
While deeper data will improve the quality of photo-zs for faint and high-z AGN, a further improvement will also come from the increasing training sample size, thanks to the Black Hole Mapper (BHM) programs at SDSS-V and 4MOST, which focus on AGN spectroscopic and on the follow-up of eROSITA X-ray sources. Together the surveys will provide spectroscopic redshifts with completeness of >60\% down to the r magnitude ~21.5 and ~22.5 for the eRASS:3 and eRASS:5, respectively, for a total of about one million sources.

\begin{acknowledgements}
      We thank the referee for their careful reading of the manuscript and the constructive feedback.
      MS thanks Dr. Marianna Lucio for the insights on features analysis and statistics.
      Part of this work was supported by the German
      \emph{Deut\-sche For\-schungs\-ge\-mein\-schaft, DFG\/} project
      Funded by \emph{Deut\-sche For\-schungs\-ge\-mein\-schaft} (DFG, German Research Foundation) under Germany's Excellence Strategy – EXC 2094 – 390783311.
    CR acknowledges support from Fondecyt Regular grant 1230345 and ANID BASAL project FB210003.
       RJA was supported by FONDECYT grant number 1231718 and the ANID BASAL project FB210003.
       
This work is based on data from eROSITA, the soft X-ray instrument aboard SRG, a joint Russian-German science mission supported by the Russian Space Agency (Roskosmos), in the interests of the Russian Academy of Sciences represented by its Space Research Institute (IKI), and the Deutsches Zentrum f\"ur Luft- und Raumfahrt (DLR). The SRG spacecraft was built by Lavochkin Association (NPOL) and its subcontractors and is operated by NPOL with support from the Max Planck Institute for Extraterrestrial Physics (MPE).

The development and construction of the eROSITA X-ray instrument were led by MPE, with contributions from the Dr. Karl Remeis Observatory Bamberg \& ECAP (FAU Erlangen-Nuernberg), the University of Hamburg Observatory, the Leibniz Institute for Astrophysics Potsdam (AIP), and the Institute for Astronomy and Astrophysics of the University of T\"ubingen, with the support of DLR and the Max Planck Society. The Argelander Institute for Astronomy of the University of Bonn and the Ludwig Maximilians Universit\"at Munich also participated in the science preparation for eROSITA.

The Legacy Surveys consist of three individual and complementary projects: the Dark Energy Camera Legacy Survey (DECaLS; Proposal ID 2014B-0404; PIs: David Schlegel and Arjun Dey), the Beijing-Arizona Sky Survey (BASS; NOAO Prop. ID 2015A-0801; PIs: Zhou Xu and Xiaohui Fan), and the Mayall z-band Legacy Survey (MzLS; Prop. ID 2016A-0453; PI: Arjun Dey). DECaLS, BASS, and MzLS together include data obtained, respectively, at the Blanco telescope, Cerro Tololo Inter-American Observatory, NSF’s NOIRLab; the Bok telescope, Steward Observatory, University of Arizona; and the Mayall telescope, Kitt Peak National Observatory, NOIRLab. Pipeline processing and analyses of the data were supported by NOIRLab and the Lawrence Berkeley National Laboratory (LBNL). The Legacy Surveys project is honored to be permitted to conduct astronomical research on Iolkam Du’ag (Kitt Peak), a mountain with particular significance to the Tohono O’odham Nation.

NOIRLab is operated by the Association of Universities for Research in Astronomy (AURA) under a cooperative agreement with the National Science Foundation. LBNL is managed by the Regents of the University of California under contract to the U.S. Department of Energy.

This project used data obtained with the Dark Energy Camera (DECam), which was constructed by the Dark Energy Survey (DES) collaboration. Funding for the DES Projects has been provided by the U.S. Department of Energy, the U.S. National Science Foundation, the Ministry of Science and Education of Spain, the Science and Technology Facilities Council of the United Kingdom, the Higher Education Funding Council for England, the National Center for Supercomputing Applications at the University of Illinois at Urbana-Champaign, the Kavli Institute of Cosmological Physics at the University of Chicago, Center for Cosmology and Astro-Particle Physics at the Ohio State University, the Mitchell Institute for Fundamental Physics and Astronomy at Texas A \& M University, Financiadora de Estudos e Projetos, Fundacao Carlos Chagas Filho de Amparo, Financiadora de Estudos e Projetos, Fundacao Carlos Chagas Filho de Amparo a Pesquisa do Estado do Rio de Janeiro, Conselho Nacional de Desenvolvimento Cientifico e Tecnologico and the Ministerio da Ciencia, Tecnologia e Inovacao, the Deutsche Forschungsgemeinschaft and the Collaborating Institutions in the Dark Energy Survey. The Collaborating Institutions are Argonne National Laboratory, the University of California at Santa Cruz, the University of Cambridge, Centro de Investigaciones Energeticas, Medioambientales y Tecnologicas-Madrid, the University of Chicago, University College London, the DES-Brazil Consortium, the University of Edinburgh, the Eidgenossische Technische Hochschule (ETH) Zurich, Fermi National Accelerator Laboratory, the University of Illinois at Urbana-Champaign, the Institut de Ciencies de l’Espai (IEEC/CSIC), the Institut de Fisica d’Altes Energies, Lawrence Berkeley National Laboratory, the Ludwig Maximilians Universitat Munchen and the associated Excellence Cluster Universe, the University of Michigan, NSF’s NOIRLab, the University of Nottingham, the Ohio State University, the University of Pennsylvania, the University of Portsmouth, SLAC National Accelerator Laboratory, Stanford University, the University of Sussex, and Texas A\&M University.

BASS is a key project of the Telescope Access Program (TAP), which has been funded by the National Astronomical Observatories of China, the Chinese Academy of Sciences (the Strategic Priority Research Program “The Emergence of Cosmological Structures” Grant \# XDB09000000), and the Special Fund for Astronomy from the Ministry of Finance. The BASS is also supported by the External Cooperation Program of Chinese Academy of Sciences (Grant \# 114A11KYSB20160057), and Chinese National Natural Science Foundation (Grant \# 12120101003, \# 11433005).

The Legacy Survey team uses data products from the Near-Earth Object Wide-field Infrared Survey Explorer (NEOWISE), a project of the Jet Propulsion Laboratory/California Institute of Technology. NEOWISE is funded by the National Aeronautics and Space Administration.

The Legacy Surveys imaging of the DESI footprint is supported by the Director, Office of Science, Office of High Energy Physics of the U.S. Department of Energy under Contract No. DE-AC02-05CH1123, by the National Energy Research Scientific Computing Center, a DOE Office of Science User Facility under the same contract, and by the U.S. National Science Foundation, Division of Astronomical Sciences under Contract No. AST-0950945 to NOAO.

Funding for the Sloan Digital Sky Survey V \footnote{\url{www.sdss.org}} has been provided by the Alfred P. Sloan Foundation, the Heising-Simons Foundation, the National Science Foundation, and the Participating Institutions. SDSS acknowledges support and resources from the Center for High-Performance Computing at the University of Utah. SDSS telescopes are located at Apache Point Observatory, funded by the Astrophysical Research Consortium and operated by New Mexico State University, and at Las Campanas Observatory, operated by the Carnegie Institution for Science.

SDSS is managed by the Astrophysical Research Consortium for the Participating Institutions of the SDSS Collaboration, including Caltech, The Carnegie Institution for Science, Chilean National Time Allocation Committee (CNTAC) ratified researchers, The Flatiron Institute, the Gotham Participation Group, Harvard University, Heidelberg University, The Johns Hopkins University, L’Ecole polytechnique f\'{e}d\'{e}rale de Lausanne (EPFL), Leibniz-Institut f\"ur Astrophysik Potsdam (AIP), Max-Planck-Institut f\"ur Astronomie (MPIA Heidelberg), Max-Planck-Institut f\"ur Extraterrestrische Physik (MPE), Nanjing University, National Astronomical Observatories of China (NAOC), New Mexico State University, The Ohio State University, Pennsylvania State University, Smithsonian Astrophysical Observatory, Space Telescope Science Institute (STScI), the Stellar Astrophysics Participation Group, Universidad Nacional Aut\'{o}noma de M\'exico, University of Arizona, University of Colorado Boulder, University of Illinois at Urbana-Champaign, University of Toronto, University of Utah, University of Virginia, Yale University, and Yunnan University.
\end{acknowledgements}

%
%
\bibliographystyle{aa} 
\bibliography{bibliography.bib}

\begin{appendix}
\section{Features Importance\label{sec:FeaturesRelevance}}
A machine learning model operating as a black box may produce predictions, but the lack of transparency renders the outcome unreliable. This not only hinders immediate interpretability but also complicates post-analyses. 
That is why it is important to understand which features are most important to our model estimates.
The ML branch investigating each feature's role in the model's predictions is commonly called "feature importance".

\subsection{Conditional Features importance\label{subsec:relevance}}
 When using the total (or model) fluxes (and colors)  in the computation of photo-z, both SED fitting and ML  techniques rely (either explicitly or implicitly) on the identification of redshift-dependent distinctive features in the SED of the sources like, for example, the Lyman and Balmer breaks. ML, however, allows us to go beyond fluxes and colors and extract information from any feature that, even if marginally, is redshift-dependent. 

With SED fitting, purely relying on the photometric bands provided by LS10, we would have used eight values (and corresponding errors) and implicitly considered 28 colors. With {\sc{CircleZ}}, we want to consider 281 features, and it is interesting to look at which features are more important, regardless of the algorithm that one may want to use to compute photo-zs. The analysis is unlike what is shown in  \citet{Dey2022}, in which the SHapley Additive exPlanations \citep[SHAP;][]{lundberg2017unified, rozemberczki2022shapley, Chen2022} was applied on photo-z computed for normal galaxies and in a specific redshift range (z<0.4), making the analysis of the result straight forward.

Our work focuses on more complex sources (i.e., AGN and QSOs) in a large redshift range (up to z=6.6). In addition, the training, test, and validation samples are distributed all over the sky, with different LS10 depths and data quality. Thus, depending on the distribution of the test and validation sources in redshift and location, the importance of the features may change from object to object. For this reason, we look at the correlation between the features and the spectroscopic redshift (thus prior to the use of {\sc{CircleZ}}) to see directly which features from LS10 are, even if marginally, redshift-dependent.  

To evaluate the strength of the relationship between the redshift and all other variables, we employed an orthogonal partial least squares \citep[OPLS][]{Trygg2002, Wehrens2007} model, applied to the validation sample. The model finds a correlation based on a linear combination of the features by projecting the predictor variables so that the resulting components explain the variation in the response variable while simultaneously separating and excluding variations orthogonal to the response variable. Before modeling, the dataset underwent preprocessing through scaling using the unit variance (UV) method. This scaling ensured that each variable contributed equally to the analysis, minimizing the potential influence of differences in variable scales.
To assess the model's performance and ensure against overfitting, we conducted cross-validation using analysis of variance (ANOVA).
ANOVA revealed statistically significant results (p-values < 0.0001), indicating no evidence of overfitting.
All the models are also validated using a 7 Folds cross validation \citep{Kohavi1995}. In such a way,  the dataset is divided into 7 equal parts, and the model is trained and validated 7 times, each time using a different part as the validation set and the remaining 6 parts for training.
The goodness of fit of the model is evaluated using the cumulative coefficient of determination, $R^2$(cum), and the goodness of prevision with $Q^2$(cum). The higher the values of $R^2$ and $Q^2$, the higher the goodness of the model that estimates the importance of the features. Table \ref{tab:Marianna}) reports the values of $R^2$ and $Q^2$ for the entire validation sample (all) and for specific subsamples: sources with z<1, z>3, point-like sources (TYPE=PSF), extended sources (TYPE$\neq$PSF). We also looked into the correlation between the features and the spectroscopic redshifts for the sources that are considered inliers and outliers after running {\sc{CircleZ}}. For this reason, we have used the validation sample on which the accuracy of {\sc{CircleZ}} has been measured.

\begin{table}[h]
\centering 
\caption{Summary table reporting the strength of the models that explains the importance of the features for the total sample and various subsamples mentioned  in Section \ref{subsec:relevance}.} 
\small
\begin{tabular}{lccc}
\hline
 Sample & $R^2$ & $Q^2$ & p-value \\
\hline 
 \hline
 \\
ALL          &  0.51  & 0.37 & 0  \\ 

z<1              & 0.43 & 0.37 & 0  \\ 

z>3         &  0.87  & 0.75 & <0.0001  \\ 

point-like             & 0.41  & 0.33 & 0  \\ 

extended             & 0.24  & 0.2 & <0.0001  \\

Inliers             & 0.60 & 0.48& 0  \\

outliers            & --  & -- & -- \\

\hline
\end{tabular}
\label{tab:Marianna}      
\end{table}

\begin{table*}[h]
    \centering
\caption{Top 15 features ranked by importance for the entire training sample and each subsample for which a correlation between redshift and the features could be found.}
    \label{tab:best15}
 
\begin{tabular}{lc||lc||lc}
\hline
\\
 All sample & value &z<1 & value &z>3 & value \\
\hline
\hline
 flux\_ivar\_w1	&3.781&		flux\_ivar\_w1 
& 2.722&		apflux\_g\_1-apflux\_z\_1	&2.217	\\
apflux\_ivar\_w1\_1	&3.478&			flux\_ivar\_w2 &	2.512	&		apflux\_g\_2-apflux\_z\_2 &	2.193 \\	
apflux\_ivar\_w1\_2	&	3.260& apflux\_ivar\_w1\_2	&	2.311&	apflux\_r\_5-apflux\_z\_5	&	3.201\\		
flux\_ivar\_w2	&	3.103	&apflux\_ivar\_w1\_3&		2.311	&			
apflux\_r\_2-apflux\_z\_2	&	2.004\\			
apflux\_ivar\_w1\_3&		2.917&apflux\_ivar\_w1\_1	&	2.239&				apflux\_r\_3-apflux\_z\_3&		2.002\\		
apflux\_g\_1-apflux\_z\_1	&	2.706		&	apflux\_ivar\_w1\_4	 &		2.235&				apflux\_g\_3-apflux\_z\_3	&	2.002 \\			
apflux\_g\_2-apflux\_z\_2&		2.677	&			apflux\_w1\_1-apflux\_w2\_1	&	2.209&				apflux\_r\_1-apflux\_z\_1	&	1.9782 \\
apflux\_ivar\_w1\_4	&	2.551	&			snr\_i	&	2.163&			snr\_g	&1.977\\		
apflux\_ivar\_w2\_1&		2.486&				snr\_w1&	2.126&				apflux\_g\_4-apflux\_z\_4	&	1.953	\\		
apflux\_g\_5-apflux\_z\_5	&	2.441&				apflux\_w1\_2-apflux\_w2\_2&	2.112	&			apflux\_r\_4-apflux\_z\_4	&	1.913 \\
apflux\_g\_3-apflux\_z\_3	&	2.336&				apflux\_ivar\_w1\_5&		2.110&			apflux\_g\_5-apflux\_z\_5&		1.885 	\\	
apflux\_ivar\_w2\_2&		2.296&				snr\_z&		2.013	&			snr\_r	&	1.8583\\				
apflux\_g\_4-apflux\_z\_4&		2.2657&				apflux\_w1\_3-apflux\_w2\_3	&	1.917&				apflux\_r\_6-apflux\_z\_6	&	1.738 \\
apflux\_ivar\_w1\_5&		2.227&				apflux\_z\_4&		1.862	&			apflux\_g\_1-apflux\_i\_1	&	1.687\\
snr\_w1&		2.175	&		apflux\_z\_5	&	1.861	&			apflux\_g\_2-apflux\_i\_2	&	1.640\\
\hline
\\
\hline
Point-like & value &Extended & value & inliers &value \\
 \hline
\hline
apflux\_g\_1-apflux\_z\_1	&3.375		&apflux\_ivar\_w1\_2&	2.461	&		flux\_ivar\_w1 &3.613\\
apflux\_g\_2-apflux\_z\_2	&3.359	&		apflux\_ivar\_w1\_1	&z,460&		apflux\_ivar\_w1\_1	&3.289	\\		
apflux\_g\_5-apflux\_z\_5	&	3.188&				flux\_ivar\_w1	&	2.455	&			apflux\_ivar\_w1\_2	&	3.087			\\
apflux\_g\_3-apflux\_z\_3	&	3.115&				flux\_ivar\_w2	&	2.433&				apflux\_g\_1-apflux\_z\_1	&	3.033		\\
apflux\_g\_4-apflux\_z\_4	&	3.058	&			apflux\_ivar\_w1\_3&		2.414&			apflux\_g\_2-apflux\_z\_2	&	3.003		\\
apflux\_g\_6-apflux\_z\_6	&	2.772&				apflux\_ivar\_w1\_4	&	2.319	&			flux\_ivar\_w2&		2.984			\\
apflux\_g\_2-apflux\_r\_2	&	2.664&				apflux\_ivar\_w2\_1	&	2.318&				apflux\_ivar\_w1\_3&		2.767			\\
apflux\_g\_7-apflux\_z\_7&		2.476&				apflux\_ivar\_w2\_2&		2.265	&			apflux\_g\_5-apflux\_z\_5	&	2.741		\\
apflux\_g\_5-apflux\_r\_5&		2.461&				apflux\_ivar\_w1\_5&		2.204&				apflux\_g\_3-apflux\_z\_3&		2.678		\\
apflux\_r\_1-apflux\_z\_1	&	2.438&				apflux\_ivar\_w2\_3&		2.164&				apflux\_g\_4-apflux\_z\_4	&	2.602			\\
flux\_ivar\_w1	&	2.385	&			apflux\_ivar\_w2\_4 &		2.029	&			apflux\_ivar\_w1\_4	&	2.417		\\
apflux\_ivar\_w1\_1	&2.293	&			snr\_w1	&	2.0165&			apflux\_g\_2-apflux\_r\_2	&	2.391			\\
apflux\_g\_1-apflux\_r\_1&		2.256	&		flux\_ivar\_r	&	1.969&				apflux\_ivar\_w2\_1&		2.342		\\
apflux\_g\_3-apflux\_r\_3&		2.242&				snr\_i	&	1.943&				apflux\_g\_6-apflux\_z\_6	&	2.288			\\
apflux\_r\_2-apflux\_z\_2	&	2.199	&			apflux\_ivar\_w2\_5&		1.884	&			apflux\_ivar\_w2\_2	&	2.173			\\
\hline
\end{tabular}
\end{table*}
\end{appendix}

The strength of the correlation between the features and the redshift is comparable between samples, with the high-z sample showing the stronger correlation (very high $R^2$ and $Q^2$) and the extended sample showing the weaker correlation. We could not find any correlation between the features and redshift for the outliers, indicating that a diversity of issues causes them. This is probably also due to the small size of the sample.
Upon completing the OPLS analysis, the model generated a ranked list of variables based on their importance in the projection. Variables with higher values in this ranking are more strongly associated with the redshift, indicating their greater influence on determining its value. This approach provides a comprehensive understanding of the relationships between variables and allows for the identification of key factors influencing the redshift prediction.\\
In all cases, no features (or groups of features) stand out for importance, and their importance decreases smoothly. This means that all features are used in some measure. Table \ref{tab:best15} reports the first 15 features for all subgroups. As mentioned at the beginning of the section, depending on redshift and subsamples, the order of importance of the features changes. For the validation sample taken globally, the first five important features are related to the error on the model flux or the smaller apertures in WISE. Then, the important features are associated with the various colors between annuli. Also, for the sample at z<1, the first important features are all related to the photometric error associated with the WISE photometry, WISE colors, and the redder of the optical bands. Considering that the bulk of the sources in the training are at low redshift, the similarity of the features importance between these two samples is not surprising.  For the sample at z>3, 13 of  15 important features are the various annuli colors. The most striking difference, however, is between the extended and point-like sources. While in the former, the higher-ranked features are related to the annuli colors, for the extended sources, the most important features are all related to the errors associated with the fluxes within various fluxes in the WISE bands. In turn, errors correlate with the photometry, and we will discuss this point further in Section \ref{sec:comparison}. One point worth noticing is that in all the cases, the features related to total fluxes appear very late in the list (usually after position 60), indicating their little impact when other features related to the light distribution are provided. Indeed, using {\sc{CircleZ}} and limiting the features to the total fluxes, their errors, and the colors provides an accuracy of 0.12 and a fraction of outliers of about 25\%.

\end{document}